\documentclass[aps,prb,twocolumn,amsfonts,amssymb,superscriptaddress,longbibliography,10pt]{revtex4-2}
\usepackage{graphicx}
\usepackage{dcolumn}
\usepackage{bm}
\usepackage{bbm,bbold}
\usepackage{color}
\usepackage{xcolor, soul}
\sethlcolor{green}
\usepackage{amsfonts}
\usepackage{amsmath}
\usepackage{mdframed}
\usepackage[normalem]{ulem}
\usepackage{mathrsfs}   
\usepackage[none]{hyphenat}
\usepackage{subfigure}
\usepackage{float}
\usepackage{cancel}
\usepackage{physics}
\usepackage{comment}
\usepackage[colorlinks=true,citecolor=blue]{hyperref}
\hypersetup{colorlinks=true,citecolor=blue,linkcolor=blue,urlcolor=blue}

\DeclareMathAlphabet\mathbfcal{OMS}{cmsy}{b}{n}

\allowdisplaybreaks 

\begin{document}

\title{Magnetic field suppression of tomographic electron transport}

\affiliation{Department of Physics, University of Bath, Claverton Down, Bath BA2 7AY, United Kingdom}
\affiliation{School of Mathematics and Statistics, The University of Melbourne, Victoria 3010, Australia}
\affiliation{Department of Engineering and Physics, Karlstad University, Karlstad, Sweden}
\affiliation{Department of Physics, Gothenburg University, 41296 Gothenburg, Sweden}
\affiliation{Nordita, Stockholm University and KTH Royal Institute of Technology, 10691 Stockholm, Sweden}

\author{Habib Rostami}
\affiliation{Department of Physics, University of Bath, Claverton Down, Bath BA2 7AY, United Kingdom}

\author{Nitay Ben-Shachar}
\affiliation{School of Mathematics and Statistics, The University of Melbourne, Victoria 3010, Australia}

\author{Sergej Moroz}
\affiliation{Department of Engineering and Physics, Karlstad University, Karlstad, Sweden}
\affiliation{Nordita, Stockholm University and KTH Royal Institute of Technology, 10691 Stockholm, Sweden}

\author{Johannes Hofmann}
\email{johannes.hofmann@physics.gu.se}
\affiliation{Department of Physics, Gothenburg University, 41296 Gothenburg, Sweden}
\affiliation{Nordita, Stockholm University and KTH Royal Institute of Technology, 10691 Stockholm, Sweden}

\date{\today}

\begin{abstract}
Degenerate two-dimensional electron liquids are theoretically established to possess two vastly distinct collisional electron mean free paths, where even-parity deformations of the Fermi surface are hydrodynamic with a short collisional mean free path but odd-parity deformations remain near ballistic (known as the ``tomographic’’ transport regime). Predicted signatures of this regime rely on the scaling of observables with temperature or device dimension, both of which are difficult to establish with certainty. Here, we consider magnetotransport in a minimal model of tomographic electrons and show that even a small magnetic field suppresses tomographic transport signatures and thus acts as a sensitive and unique probe of this regime. Fundamentally, the magnetic field breaks time-reversal invariance, which is a prerequisite for the odd-even parity effect in the collisional relaxation. We analyze in detail the scaling of the transverse conductivity, which has been linked to small-channel conductance of interaction-dominated electrons, and show that a tomographic scaling regime at intermediate wave numbers is quickly suppressed with magnetic field to a hydrodynamic or collisionless form. We confirm that the suppression occurs at relatively small magnetic fields when the cyclotron radius is comparable to the ballistic mean free path of the dominant odd-parity mode. This occurs at a much smaller magnetic field than the magnetic field strength required to suppress hydrodynamic electron transport, which suggests an experimental protocol to extract the odd-parity mean free path.
\end{abstract}

\maketitle

\section{Introduction}\label{sec:intro}

Recent advances in the fabrication of ultraclean materials allow to probe interaction-dominated electron transport, for which the electron mean free path due to binary collisions is smaller than the device dimension or the phonon or impurity mean free paths~\cite{fritz24}. This leads to collective hydrodynamic electron flow with a characteristic scale set by the electron shear viscosity. Hydrodynamic transport is  expected to be the generic mode of transport in clean materials, and signatures of hydrodynamic electrons have by now been reported in a wide variety of (predominately two-dimensional) materials like monolayer~\cite{crossno16,bandurin16} and bilayer graphene~\cite{nam17,bandurin18}, (Al,Ga)As~\cite{dejong95,buhmann02,ginzburg21}, WTe$_2$~\cite{vool21,aharonsteinberg22}, WP$_2$~\cite{gooth18}, PdCoO$_2$~\cite{moll16}, MoP~\cite{kumar19}, and NbP~\cite{gooth17}. Since by Fermi liquid theory, electron interactions are phase-space suppressed at low temperature, the hydrodynamic regime is expected to span an intermediate temperature range where temperatures are large enough that electron interactions dominate over impurity scattering yet high-temperature relaxation due to phonon scattering is not yet prevalent. 

In this context, it has been theoretically established that the Pauli principle imposes very different phase space restrictions on electron relaxation depending on the parity of the Fermi surface perturbation~\cite{gurzhi95,ledwith19,ledwith17,hofmann23,nilsson24}, which leads to the emergence of not one but two vastly different electronic mean free paths: First, even-parity deformations of the Fermi distribution, which are described by a momentum-parity symmetrized distribution \mbox{$f_e({\bf k}) = [f({\bf k}) + f(-{\bf k})]/2$}, relax with a short mean free path \mbox{$\ell_e$} that at low temperatures increases as an inverse square of temperature, \mbox{$\ell_e \sim v_F T_F/T^2$} (where $v_F$ is the Fermi velocity and $T_F$ the Fermi temperature)~\cite{pines18,baym91,giuliani05}. Second, by contrast, odd-parity deformations, described by the anti-symmetrized distribution \mbox{$f_o({\bf k}) = [f({\bf k}) - f(-{\bf k})]/2$}, have a much larger anomalously increased ballistic mean free path, \mbox{$\ell_o \gg \ell_e$}. Indeed, for a simple circular Fermi surface (applicable to, for example, GaAs or doped graphene), more recent studies predict an asymptotic low-temperature scaling of \mbox{$\ell_o \sim v_F T_F^3/(m^4T^4)$}, which is significantly increased at low temperatures, with an additional dependence on an angular-momentum index~$m$ that labels different odd-parity modes~\cite{ledwith19,ledwith17,hofmann23,nilsson24}. This odd-even effect in the quasiparticle relaxation is surprisingly robust and predicted to exist up to temperatures \mbox{$T\lesssim0.1T_F$} with a finite number of \mbox{${\it O}(\sqrt{T_F/T})$} decoupled odd-parity modes~\cite{ledwith19,ledwith17,hofmann23,nilsson24}. Quite generally, the odd-even effect will be present for any system with a parity-even Fermi surface, for which deformations can be classified into even-parity and odd-parity modes. Fundamentally, the symmetry that guarantees this is time-reversal invariance, which implies a parity-even electron dispersion, \mbox{$\varepsilon({\bf k}) = \varepsilon(-{\bf k})$}~\cite{girvin19}.

Electron flows exhibiting the odd-even effect have been dubbed ``tomographic'' to distinguish this new regime from hydrodynamic and ballistic flows. Recent predictions for phenomena exhibited by these flows include an anomalous fractional scaling of a narrow channel's conductivity with its width~\citep{ledwith19b}, distinct bulk collective modes~\citep{hofmann22}, and an enhanced nonlinear response~\cite{hofmann23b}. Indeed, two recent experiments report signatures that could be attributable to tomographic flow: First, the observed Hall field profile in a narrow channel is seen to have a slightly larger curvature compared to the strict hydrodynamic Poiseuille profile~\cite{sulpizio19}, an observation that appears not fully explained so far but in Ref.~\cite{sulpizio19} is speculated to be due to odd-parity modes. Second, the viscosity as extracted from a hydrodynamic modeling of electron flow in a Corbino geometry shows an anomalous inverse linear-in-temperature scaling at low temperatures~\cite{zeng24}. This result is unanticipated since the hydrodynamic shear viscosity involves a microscopic even-parity deformation with a relaxation time that (according to Fermi liquid theory, see the above discussion) increases with the inverse square of the temperature~\cite{steinberg58,gran23}. Quite generally, a challenge in detecting existing predictions for tomographic phenomena is that establishing an anomalous scaling with device dimensions requires the fabrication (and theoretical modeling) of different samples, and an anomalous temperature scaling must be distinguished from other relaxation due to phonons and impurities. It would therefore be desirable to have an in-situ method to suppress tomographic effects without changing samples or parameter values. This would discriminate tomographic effects from other nonhydrodynamic phenomena. 

In this paper, we include magnetic field effects on tomographic transport and demonstrate that due to the breaking of time-reversal invariance, a tomographic scaling regime is rapidly suppressed with increasing magnetic fields (illustrated in Figs.~\ref{fig:1} and~\ref{fig:2} below). This implies that if anomalous transport observations are indeed due to tomographic flow, then a hallmark signature would be a rapid suppression of such signatures with magnetic fields to a more conventional hydrodynamic or collisionless form (with a subsequent suppression to standard Ohmic transport at even larger fields). Here, we focus in our analysis on the transverse static conductivity, which is linked to the the shear viscosity and which addresses the recent experiment~\cite{zeng24}. However, the general effect of a magnetic-field induced suppression of tomographic transport should hold more broadly~\cite{gurzhi95}. In particular, as an observable signature in magnetoresistance measurements~\cite{zeng24}, this suggest a minimum of the magnetoresistance factor at intermediate magnetic fields.

\begin{figure}
    \centering
    \includegraphics{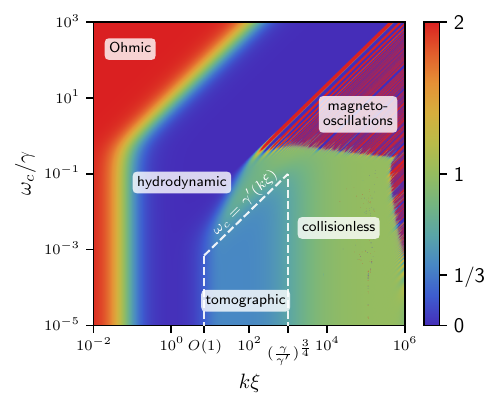}
    \caption{Density plot of the logarithmic derivative of \mbox{$k^2\sigma_T$} with respect to \mbox{$k\xi$} (with \mbox{$\xi=v_F/\sqrt{\gamma\gamma'}$}) as a function of \mbox{$k\xi$} and \mbox{$\omega_c/\gamma$} on a logarithmic scale. The parameter values for the odd-mode damping are \mbox{$\gamma'/\gamma = 10^{-4}$}  and for impurity scattering \mbox{$\gamma_i/\gamma = 10^{-7}$}. The logarithmic derivative determines the local power-law scaling exponent $\beta$ in \mbox{$k^2\sigma_T \sim (k\xi)^\beta$}, which reveals distinct hydrodynamic (\mbox{$\beta = 0$}, dark blue), tomographic (\mbox{$\beta=1/3$}, light blue),  collisionless (\mbox{$\beta=1$}, green), and Ohmic (\mbox{$\beta = 2$}, red) transport regimes.}
    \label{fig:1}
\end{figure}

\begin{figure}[b!]
\includegraphics{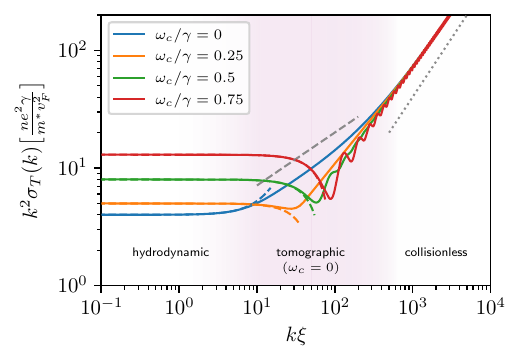}
\caption{
Transverse conductivity $\sigma_T$ (multiplied by $k^2$) as a function of rescaled wave number $k\xi$  with \mbox{$\xi=v_F/\sqrt{\gamma\gamma'}$} and \mbox{$\gamma' = 0.001 \gamma$} for different values of the cyclotron frequency $\omega_c$ measured in units of $\gamma$, \mbox{$\omega_c/\gamma = 0,0.25,0.5,$} and $0.75$ (blue, yellow, green, and red curves, respectively). Without magnetic field (blue line), we distinguish a hydrodynamic regime with \mbox{$k^2\sigma \sim (k\xi)^0$}, a tomographic regime (pink shaded region) with \mbox{$k^2\sigma \sim (k\xi)^{1/3}$} (gray dashed line), and a a collisionless regime with \mbox{$k^2\sigma \sim (k\xi)^1$} (gray dotted line). Dashed lines in matching colors indicate the analytical limits from the derivative expansion, Eq.~\eqref{eq:sigmaTexpansion}. At finite magnetic fields, the intermediate tomographic regime is suppressed.
}
\label{fig:2}
\end{figure}

The present study is motivated by the expectation that the magnetic field competes with all possible relaxation processes. Specifically, based only on quasiclassical arguments one can argue that as soon as the cyclotron radius \mbox{$r_c = m^* v_F/eB$} (here $m^*$ is the effective electron mass and \mbox{$e>0$} the magnitude of the electron charge) becomes smaller than a mean free path scale of interest~$\ell$, the corresponding relaxation-related effects must be suppressed. The characteristic magnetic field scale is thus \mbox{$B_\star \simeq m^* v_F/e \ell$}. Introducing the cyclotron frequency \mbox{$\omega_c = eB/m^*$}, we find that the relaxation channel of interest is suppressed after the cyclotron frequency reaches a corresponding  inverse relaxation time scale, i.e., the decay rate \mbox{$v_F/\ell$}. In clean Fermi liquids, we have \mbox{$\ell_{\rm min}\sim \ell_e$}, and therefore all electronic relaxation effects should be suppressed for \mbox{$B_\star \simeq m^* v_F/e \ell_e$}, an effect that is elegantly utilized in the experiment by~\textcite{zeng24} to disentangle Ohmic and electron hydrodynamic transport contributions. We will argue in this paper that the characteristic magnetic field that impedes the tomographic regime is much smaller than the scale at which hydrodynamic transport is suppressed. Indeed, we find that it is proportional to the inverse mean free path of the dominant odd-parity mode at the corresponding wave number.

\begin{figure}[t!]
\includegraphics{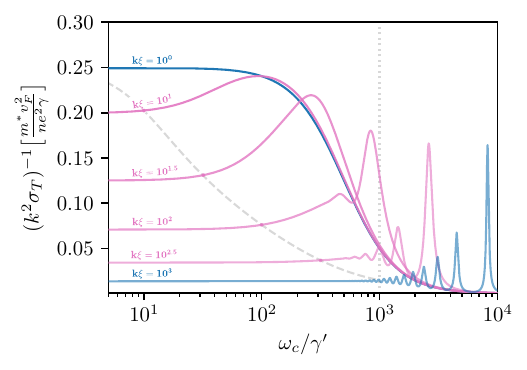}
\caption{
Inverse of the transverse conductivity $\sigma_T$ (multiplied by $k^2$) with \mbox{$\gamma'/\gamma = 10^{-3}$} as a function of rescaled cyclotron frequency $\omega_c/\gamma'$ at different wave numbers (top to bottom) $k\xi = 10^{0},10^1,10^{1.5},10^2,10^{2.5}$, and $10^3$. Curves with conductivities starting in the tomographic regime at zero field (which exists for a momentum range \mbox{$1 \lesssim k\xi \lesssim (\gamma/\gamma')^{3/4}$, cf.~Fig.~\ref{fig:2}}) are shown in pink, and in blue otherwise. Deviations from tomographic scaling start at (weak) field strengths \mbox{$\omega_c \simeq \gamma' (k\xi)$} (gray dashed line), where the magnetic cyclotron length is comparable to the mean free path of the dominant odd-parity mode at this wave number. This suppression of tomographic scaling occurs at a much smaller cyclotron frequency than the even-parity damping $\gamma$ (gray dotted line).
}
\label{fig:3}
\end{figure}
\begin{figure*}
\includegraphics{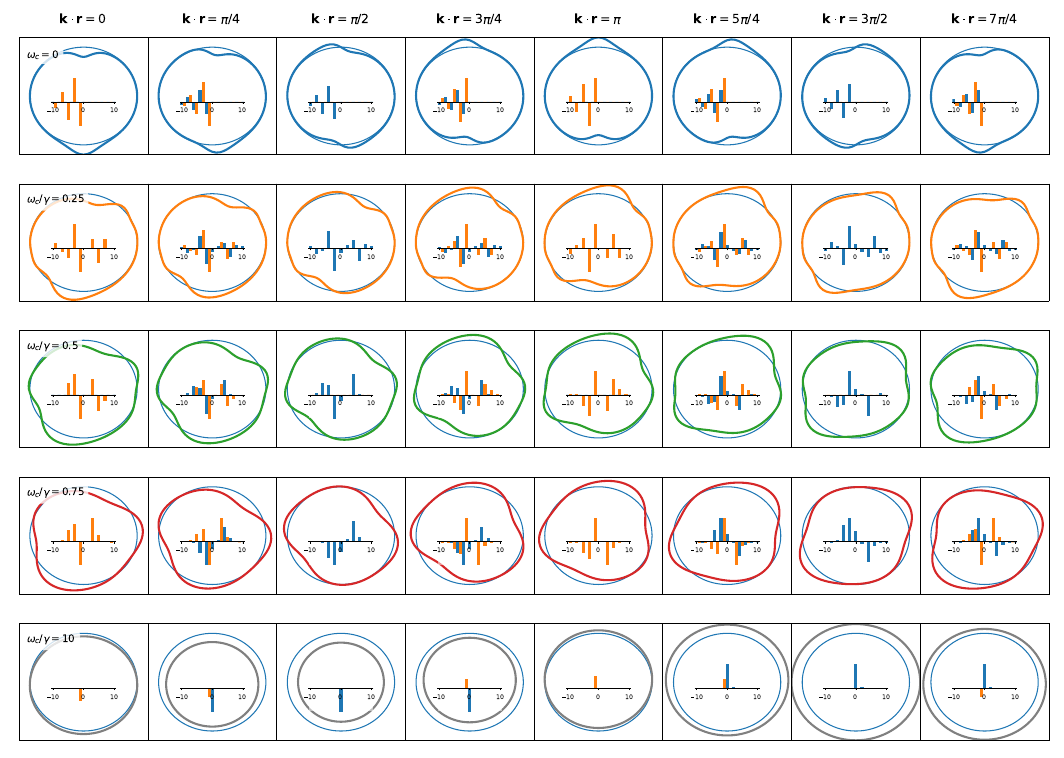}
\caption{
Fermi surface deformations in response to a static perturbation \mbox{$\mathbf{E}=E_y \cos({\bf k} \cdot {\bf r})$} at different points in space (left to right panels). The external momentum in all plots is $k\xi = 100$. We start in the tomographic regime at $\omega_c=0$ and gradually increase the magnetic field (top to bottom panels). The parameters and color coding are the same as in Fig.~\ref{fig:2}. The insets show a histogram of the Fourier coefficients $a_n$ and $b_n$ of the Fermi surface deformation, Eq.~\eqref{eq:fourierseries}, where the $a_n$ are on the positive axis and the $b_n$ on the negative axis, and even-parity coefficients are blue and odd-parity coefficients are orange.
 }
\label{fig:4}
\end{figure*}

The focus of this study is to establish and illustrate the principles behind the magnetic-field induced suppression of tomographic effects. We therefore adopt a minimal model of tomographic electrons that allows to make semi-analytical calculations. In this model, we assume rigid deviations of the Fermi surface (as opposed to a temperature or interaction-induced broadening of the Fermi edge) and parametrize deformations of the quasiparticle distribution from local equilibrium as
\begin{align}
\delta f( {\bf r}, {\bf p}) &= e^{i {\bf k} \cdot {\bf r}} \biggl(- \frac{\partial f_0}{\partial \varepsilon}\biggr) h(\theta) , \label{eq:deformation}
\end{align}
where ${\bf k}$ is the wave vector of the perturbation (we assume static perturbations in the following), and we separate a nonanalytical  factor involving the derivative of the equilibrium Fermi-Dirac distribution $f_0$. In this way, by the chain rule, $h(\theta)$ can be thought of a rigid deformation of the chemical potential that depends on the angle $\theta$ of the momentum vector under the Fermi surface  (see, for example, Fig.~\ref{fig:4} below for illustrations). This function can be expanded in angular harmonics,
\begin{align}
h(\theta) &= \sum_{m=-\infty}^\infty e^{i m \theta} h_m , \label{eq:hangular}
\end{align}
with an angular momentum index $m$. The even- and odd-parity components of the distribution function discussed above are then composed of even and odd~$m$, respectively. For a circular Fermi surface, binary collisions will not change the angular momentum index, such that collisions can be treated in a generalized relaxation-time approximation with individual decay rates \mbox{$\gamma_m \sim v_F/\ell_m$} that are inversely proportional to the mean free path $\ell_m$ of the mode. Here, the first modes \mbox{$m=0,\pm1$} (the ``hydrodynamic'' modes) describe a compression and a rigid center-of-mass shift of the Fermi surface, respectively, which have infinite lifetime since they are linked to conserved quantities (binary collision do not change particle number or total momentum), i.e., \mbox{$\gamma_0=\gamma_{\pm 1}=0$}. Moreover, we will assume that all remaining even-parity modes follow standard Fermi-liquid scaling, \mbox{$\gamma_{m \, {\rm even}} = \gamma$}, with no pronounced $m$-dependence. For odd-parity modes, we adopt a Matthiessen rule (with \mbox{$\gamma'\ll \gamma$}),
\begin{align}
\gamma_{m \, {\rm odd}} = \frac{1}{\frac{1}{\gamma} + \frac{1}{\gamma' m^4}} , \label{eq:odddamping}
\end{align}
that interpolates between the relaxation rate of the lowest $m$ modes with anomalously long mean free path and $m$-independent scaling for large $m$. We follow Fermi liquid conventions and do not assume a strong dependence of the relaxation rates on the magnetic field~\cite{lifshitz81}. The form~\eqref{eq:odddamping} is an accurate description of the lowest odd eigenvalues at low temperatures and is in good agreement with exact diagonalization studies of the Fermi liquid collision integral~\cite{hofmann23,nilsson24}. Note that elastic impurity scattering (with scattering rate $\gamma_i$) will contribute to all angular modes except for \mbox{$m=0$}~\cite{hofmann22}. The advantage of the parametrization~\eqref{eq:deformation} is that the kinetic Fermi liquid equation takes an effective tight-binding form (omitting the Landau parameters here),
\begin{align}
& \bigl(\gamma_m - i m \omega_c\bigr) \ h_m + i \frac{v_F k}{2} ( h_{m+1} +  h_{m-1}) \nonumber \\[1ex]
&\quad = \frac{v_F e E}{2} \bigl(e^{i\theta_E} \delta_{m,1} + e^{-i\theta_E} \delta_{m,-1}\bigr) , \label{eq:tightbindingintro}
\end{align}
where we include an external electric field ${\bf E}$ that forms an angle $\theta_E$ with the wave vector ${\bf k}$ [see Appendix~\ref{sec:introkinetic} for a derivation of the kinetic equation in a magnetic field, and Appendix~\ref{sec:harmonicprojection} for a derivation of the tight-binding representation~\eqref{eq:tightbindingintro}]. The left-hand side of Eq.~\eqref{eq:odddamping} is reminiscent of a one-dimensional tight-binding Hamiltonian with lattice sites labeled by $m$, where the term $v_F k/2$ fixes the amplitude of nearest-neighbor hopping, the magnetic field corresponds to an effective constant electric field, and $\gamma_m$ is a non-Hermitian on-site decay term. For the tight-binding representation~\eqref{eq:tightbindingintro}, very efficient exact numerical solution algorithms exist to solve for the vector $\{h_m\}$~\cite{hofmann22,setiawan22}, which do not rely on a slowly convergent matrix inversion of Eq.~\eqref{eq:tightbindingintro} with finite cutoff, and which are used to obtain the numerical solutions in this paper.

This paper is structured as follows: In Sec.~\ref{sec:mainresults}, we summarize the main results of our analysis and discuss the magnetic field dependence of the transverse conductivity. Section~\ref{sec:kinetic} presents a detailed discussion of the minimal kinetic model for odd-mode transport outlined above and describes the crossover between the hydrodynamic, tomographic, and collisionless regimes. We conclude the paper with a summary and outlook in Sec.~\ref{sec:outlook}. Basics of Fermi liquid theory in a magnetic field, details of the evaluation of the kinetic equation at large momenta, as well as a link between a derivative expansion and Hilbert expansions used in hydrodynamics are relegated to five Appendices. 

\section{Summary of main results}\label{sec:mainresults}

We begin by summarizing the main results of this paper for the magnetic field dependence of the (static) transverse conductivity~$\sigma_T$. Previous discussions of the transverse conductivity without an applied magnetic field in the context of an odd-even effect are Refs.~\cite{ledwith19,hofmann22,kryhin23}. The conductivity at finite wave vectors is often taken a measure of the conductance in narrow channels, where the channel width~$W$ sets a characteristic wave number \mbox{$k \simeq 1/W$}. While such an identification neglects phenomena arising from an accurate modeling of the scattering at channel edges (e.g., boundary layers and velocity slip), it includes finite-wavelength effects, and is the basis of the analysis of the experiment~\cite{zeng24}, where an effective viscosity in the Corbino geometry is identified with the inverse conductivity in a clean system, \mbox{$\eta_{\rm eff} \sim n^2e^2/(k^2\sigma_T)$}~\cite{forster18}. For an extended discussion of the  conductance for tomographic flow in a channel that includes slip boundary conditions and the boundary layer, see Ref.~\cite{benshachar25}.

Formally, within linear response, the the conductivity tensor links the electric charge current $j_i$ to the electric field through  \mbox{$j_i(\mathbf{k}) = \sigma_{ij}(\mathbf{k}) E_j(\mathbf{k})$}, which is decomposed as \mbox{$\sigma_{ij} = \frac{k_i k_j}{k^2} \sigma_L + \bigl(\delta_{ij} - \frac{k_i k_j}{k^2}\bigr) \sigma_T + \varepsilon_{ij} \sigma_H$}, with $\sigma_L$, $\sigma_T$, and $\sigma_H$ the longitudinal, transverse, and Hall conductivity components, respectively. For definiteness, we pick ${\bf E}$ along the $y$ and ${\bf k}$ along the $x$ direction, 
\begin{align} 
{\bf E}(t) = {\rm Re} \, \{ e^{i k x} \, E \} \hat{\bf e}_y , \label{eq:planewave}
\end{align} 
such that the transverse conductivity follows from the $y$ component of the current, \mbox{$j_y = \sigma_T E$}, and the $x$ component sets the Hall response, \mbox{$j_x = \sigma_H E$}. 

To convey our key message and to summarize the results discussed in this paper in a unified way, we show in Fig.~\ref{fig:1} the different scaling regimes of the scaled transverse conductivity \mbox{$k^2\sigma_T$} as a function of both wave number (measured by the dimensionless variable $k\xi$) and magnetic field (measured by the cyclotron frequency $\omega_c$), where the characteristic length scale 
\begin{align}
\xi=\frac{v_F}{\sqrt{\gamma \gamma'}}
\end{align}
is set by the geometric mean of the even and odd decay rate scales $\gamma$ and $\gamma'$, respectively. 
Figure~\ref{fig:1} shows a density plot of the logarithmic derivative of the transverse conductivity with respect to the parameter $k\xi$,
\begin{align}
    \beta(k\xi,\omega_c) &= \frac{\partial \ln (k^2\sigma_T)}{\partial \ln (k\xi)} ,
\end{align}
which defines a local scaling exponent with $k\xi$, \mbox{$k^2\sigma_T \sim (k\xi)^\beta$}. This scaling exponent is linked to a characteristic dependence of the channel conductance~\mbox{$G\sim W^{3-\beta}$} on the channel width $W$~\cite{ledwith19b}. Given that in a clean translation-invariant metal that we investigate here, the static transverse conductivity has a hydrodynamic \mbox{$k^{-2}$} singularity as \mbox{$k\to 0$}, we resolve the singularity and show results for the conductivity \mbox{$\sigma_T(k)$} multiplied by $k^2$ throughout the paper.  In the figure, we use an odd-mode damping of \mbox{$\gamma'/\gamma = 10^{-4}$}, and for full generality, we also include an impurity relaxation with strength \mbox{$\gamma_i/\gamma = 10^{-7}$}.

In the Ohmic regime (red) at small \mbox{$k\xi$} and large $\omega_c$, the conductivity is constant, \mbox{$\sigma_T = ne^2/(m^*\gamma_i)$}, and hence \mbox{$\beta=2$}. At larger momenta, where interactions dominate over impurity scattering, the Ohmic regime gives way to the hydrodynamic regime (blue). Here, the hydrodynamic divergence of the conductivity implies a constant \mbox{$k^2\sigma_T$} and hence \mbox{$\beta=0$} (cf.~also Fig.~\ref{fig:2} and Eq.~\eqref{eq:sigmaTexpansion} below). Increasing the momentum even further, this is followed by a transition to a collisionless regime (green), which is direct at larger magnetic fields, and in which the conductivity is a linear function of $k\xi$, \mbox{$\beta=1$} [cf.~Fig.~\ref{fig:2} and Eq.~\eqref{eq:sigmacolless} below]. For large momenta and magnetic fields, the collisionless linear-in-$k\xi$ scaling is superimposed by magneto-oscillations (red/blue striped area) in the conductivity [cf.~Fig.~\ref{fig:2} and Eq.~\eqref{eq:asymptoticmagnetic} below]. Finally, and most importantly, at smaller magnetic fields we find the separate tomographic regime (cyan) in between the hydrodynamic and collisionless regions, where \mbox{$\beta=1/3$} [cf.~also Fig.~\ref{fig:2} and Eq.~\eqref{eq:sigmatomographic} below]. This regime is restricted to an intermediate momentum regime \mbox{$1\lesssim k\xi \leq (\gamma/\gamma')^{3/4}$} (vertical dashed lines). Seen as a function of magnetic field, the tomographic region is restricted to a wedge that is terminated by the (small) critical magnetic field (dashed line in Fig.~\ref{fig:1}) 
\begin{align}
    \omega_c^{\rm supp} \simeq \gamma' (k\xi) \ll \gamma , \label{eq:criticalfield}
\end{align}
which corresponds to a cyclotron radius of \mbox{$kr_c \simeq \sqrt{\gamma'/\gamma}$} that matches the dominant odd-parity mean free path (cf.~Fig.~\ref{fig:3} and Sec.~\ref{sec:deformation}). This suppression of the tomographic transport window with magnetic field is the key result of our study.

\subsection{Scaling at fixed magnetic fields}

In more detail, we plot in Fig.~\ref{fig:2} the transverse conductivity as a function of the rescaled wave number~$k\xi$ and fixed magnetic field (corresponding to horizontal cuts in Fig.~\ref{fig:1}). In the absence of the magnetic field (\mbox{$\omega_c=0$}, blue curve), the low-momentum hydrodynamic and the high-momentum collisionless regimes are clearly separated by an intermediate tomographic regime (indicated by the pink shaded region) with a characteristic power-law momentum scaling \mbox{$k^2 \sigma_T\sim (k\xi)^{1/3}$}~\cite{ledwith19b} (indicated by the dashed gray line). The hydrodynamic-tomographic and tomographic-collisionless transitions occur for wave numbers \mbox{$k\xi  = {\it O}(1)$} and \mbox{$k\xi\sim (\gamma/\gamma')^{3/4}$}, respectively. If we assume the asymptotic low-temperature scaling of the odd- and even-mode damping rates discussed in the Introduction (where the even-parity decay rate follows a standard Fermi-liquid form \mbox{$\gamma \simeq (T/T_F)^2$} and the odd-parity rates are suppressed by two further powers of temperature \mbox{$\gamma' \simeq (T/T_F)^4$}), this gives an anomalous temperature scaling \mbox{$k^2 \sigma_T \sim T$}~\cite{kryhin23}. Identifying an effective shear viscosity by~\mbox{$\eta_{\rm eff} = n^2e^2/(k^2\sigma_T)$}, this temperature scaling is consistent with the recent experimental observation~\cite{zeng24} and distinct from the expected hydrodynamic Fermi-liquid scaling \mbox{$\eta \sim 1/\gamma \sim 1/T^2$}~\cite{steinberg58,gran23}. While it would be interesting to discuss the regime of validity of the tomographic analysis as a function of temperature and device parameter values, here we focus on investigating the effects of an external magnetic field on this scaling.

Figure~\ref{fig:2} also shows the transverse conductivity for three magnetic fields with moderate strength compared to the even-parity damping rate of $\omega_c/\gamma =0.25,0.5,$ and $0.75$ (orange, green, and red lines, respectively). It is clear from the figure that all these magnetic fields are sufficiently large to destroy the tomographic scaling. At small $k\xi$, the transverse conductivity to second order in \mbox{$k\xi$} follows from a derivative expansion (described in detail in Sec.~\ref{sec:kinetic}), which reads
\begin{align}
    k^2 \sigma_T &= \frac{ n e^2\gamma}{m^* v_F^2} \biggl\{ 4 \biggl(1 + \frac{4\omega_c^2}{\gamma^2}\biggr) \nonumber \\[1ex]
    &\quad + \frac{(v_Fk/\gamma)^2}{1 + (3 \omega_c/\gamma_3)^2} \biggl(\frac{\gamma}{\gamma_3} - \frac{4 \omega_c^2}{\gamma_3^2} \Bigl(3+ \frac{\gamma_3}{\gamma}\Bigr)\biggr) \biggr\} \nonumber \\[1.5ex]&\quad +{\it O}\left((k\xi)^4\right) . \label{eq:sigmaTexpansion}
\end{align}
The first term in braces in Eq.~\eqref{eq:sigmaTexpansion} is the standard hydrodynamic result with a magnetic-field induced enhancement of the conductivity~\cite{steinberg58,alekseev16}. Compared to the zero-field case, this limit remains valid up to much larger wave numbers (by about one order in magnitude for the parameter values in Fig.~\ref{fig:2}, to about \mbox{$k\xi \simeq 10^1$}). The second term in braces in Eq.~\eqref{eq:sigmaTexpansion} gives the leading-order finite-wavelength correction to the hydrodynamic conductivity. This correction depends only on the lowest damping rates (specifically, the damping $\gamma_2$ of the quadrupole \mbox{$m=2$} mode and the damping $\gamma_3$ of the first odd \mbox{$m=3$} mode with trigonal symmetry). The latter dependence makes it sensitive to the tomographic effect. Equation~\eqref{eq:sigmaTexpansion} is indicated by dashed lines with matching colors in Fig.~\ref{fig:2} and is in good agreement with the exact solution at long wavelengths. As we will show, the asymptotic solution in Eq.~\eqref{eq:sigmaTexpansion} starts to deviate from the exact solution once higher $m$ harmonics participate in the current response. The leading-order tomographic correction in Eq.~\eqref{eq:sigmaTexpansion} changes sign at a moderate magnetic field strength
\begin{align}
\omega_c^\star &= \frac{\gamma}{2} \sqrt{\frac{\gamma_3}{3 \gamma + \gamma_3}} \approx \frac{3}{2} \sqrt{3 \gamma \gamma'} ,
\end{align} 
with a scale that in the presence of the odd-even effect is set by the geometric mean of the even- and odd-mode damping. Interestingly, for \mbox{$\omega_c>\omega_c^\star$}, the leading correction to the conductivity is negative, i.e.,  opposite to that at zero field, with a reduction in the conductivity instead of an enhancement. At large wave numbers, a magnetic-dominated regime emerges, which exhibits oscillations as a function of the cyclotron frequency. Here, the conductivity follows a linear-in-momentum scaling as in the collisionless limit with an additional superimposed oscillatory form. These can be understood via a judiciously chosen basis transformation for the conductivity expectation value~\cite{lee75,simon93}, reviewed in Appendix~\ref{app:basis_high}, which gives
\begin{align}
k^2 \sigma_T &= \frac{n e^2\gamma}{m^* v_F^2} \times 2 \omega_c \biggl[ X \biggr(\coth(\pi R) - \frac{\sin(2X)}{\sinh(\pi R)}\biggr) \nonumber \\[1ex]
&\qquad - \biggr(R - \Bigl(R^2-\frac{3}{4}\Bigr) \frac{\cos(2X)}{\sinh(\pi R)}\biggr)\biggr] , \label{eq:asymptoticmagnetic}
\end{align}
where we define \mbox{$X=v_Fk/\omega_c$} and \mbox{$R=\gamma/\omega_c$}. This expression has an oscillation period \mbox{$\Delta k = \omega_c/v_F$}. The linear-in-momentum conductivity prediction of Eq.~\eqref{eq:asymptoticmagnetic} is shown in Fig.~\ref{fig:2} with a dotted line. Crucially, Fig.~\ref{fig:2} shows that no intermediate tomographic region is present at finite magnetic field, and there is no  independent anomalous power-law scaling of the conductivity.

\subsection{Scaling at fixed wave number}

We corroborate this picture in Fig.~\ref{fig:3}, which shows the transverse conductivity as a function of magnetic field at fixed magnetic wave number (corresponding to vertical cuts in Fig.~\ref{fig:1}), and which illustrates the competition between relaxation effects and the magnetic field in transport. Note that compared to Fig.~\ref{fig:1}, the figure shows the scaled transverse conductivity versus the cyclotron frequency measured in units of $\gamma'$, the smallest relaxation frequency scale. For different wave numbers, we observe deviations from the tomographic behavior at cyclotron frequencies \mbox{$\omega_c \simeq \gamma' (k\xi) \ll \gamma$} that compete with the dominant odd-parity decay mode at a given wave number. In the figure, we indicate curves with momenta that are in the tomographic regime at zero field in pink, and we mark the critical field~\eqref{eq:criticalfield} at which the tomographic limit is suppressed by the dashed gray line. The point of deviation agrees very accurately with the prediction~\eqref{eq:criticalfield}. Beyond this point, the microscopic Fermi surface deformation evolves smoothly and is increasingly suppressed at high fields as discussed above. At large magnetic fields, the Fermi liquid is in a collisionless regime and, as is apparent in Fig.~\ref{fig:3}, displays strong magneto-oscillations that are eventually suppressed. It is worth noting that these oscillations with sharp resonant peaks are known as classical magnetic oscillations, which differ from quantum oscillations related to Landau levels~\cite{dmitriev12,hofmann19}.

\subsection{Fermi surface deformation}\label{sec:deformation}

To determine the critical strength at which the magnetic field suppresses tomographic transport beyond the small-momentum limit, it is instructive to analyze the full Fermi surface deformation in the tomographic regime and the effect of a background magnetic field. In Fig.~\ref{fig:4}, we illustrate the local Fermi surface deformation for the magnetic-field parameters chosen in Fig.~\ref{fig:2} (top row to bottom row), where each subfigure shows
\begin{align}
    h(\theta, x) = {\rm Re} \{e^{ikx} h(\theta)\} 
\end{align}
at equidistant positions \mbox{$kx = 0-2\pi$} (left to right panels) as determined from a numerical solution of the Fermi-liquid equations~\eqref{eq:tightbindingintro} in response to the transverse periodic electric field~\eqref{eq:planewave}. The insets show the angular Fourier spectrum, i.e., the real-valued coefficients in the expansion
\begin{align}
    h(\theta, x) = \sum_{n\geq 0} a_n \cos n\theta + \sum_{n>0} b_n \sin n \theta , \label{eq:fourierseries}
\end{align}
with $a_n$ on the positive axis and $b_n$ on the negative axis, and orange and blue denoting odd and even components, respectively. In particular, the component~$b_1$ gives the transverse current, and the component~$a_1$ the Hall current. This expansion is related to the complex exponential expansion in Eq.~\eqref{eq:hangular} via \mbox{$a_n=(h_n-h_{-n})/(2i)$} and \mbox{$b_n=(h_n+h_{-n})/2$}. 

In the zero-field case, the Fermi liquid forms stripes of width \mbox{$\Delta x = \pi/k$} that carry a transverse current (i.e., a current in the $y$ direction) with alternating direction. At the point of maximal current (blue curves for \mbox{$kx = 0,\pi$} in Fig.~\ref{fig:4}), the Fermi surface deformation has odd parity and is well described by a variational form that is exponentially localized around the point \mbox{$\theta = \pi/2$},
\begin{align}
    h_{\rm tomo}(\theta) &\sim \biggl(\frac{8\pi}{\delta \theta^2}\biggr)^{1/4} e^{-(\theta -\tfrac{\pi}{2})^2/\delta \theta^2}, \label{eq:trialtomographic}
\end{align}
and correspondingly (with reversed sign) for the lower peak \mbox{$\theta = -\pi/2$}. The curves in Fig.~\ref{fig:4} are normalized such that the deviation in each plot is at most $20\%$ of the Fermi surface. As will be discussed in Sec.~\ref{sec:kinetic}, this deformation is optimal for an angular spread of
\begin{align}
    \delta \theta &= \biggl(\frac{2 \sqrt{15}}{k\xi}\biggr)^{1/3} , \label{eq:widthtomographic}
\end{align}
which becomes increasingly narrow at large momenta. The even-parity component, which describes the Fermi surface deformation at the points \mbox{$kx = \pi/2,3\pi/2$} follows from~\cite{ledwith17}
\begin{align}
    h_{\rm even}(\theta) &= - \frac{i {\bf k} \cdot {\bf v}}{\gamma} h_{\rm tomo}(\theta) , \label{eq:oddevenrelation}
\end{align}
and describes the interface between stripes with opposite local current flow. 

At nonzero magnetic fields, the general structure of the Fermi surface deformation remains, with two main differences: First, the time-reversal-breaking magnetic background mixes transverse and longitudinal Fourier harmonics $a_n$ and $b_n$ (see, for example, the column with \mbox{$\mathbf{k}\cdot \mathbf{r}=0$}), such that the simple variational form~\eqref{eq:trialtomographic} no longer applies. Second, as the magnetic field increases, the higher angular-momentum modes are suppressed, implying a broader Fermi surface deformation. Such an effect is expected from the tight-binding representation~\eqref{eq:tightbindingintro}, where the magnetic field corresponds to an effective constant electric field that increases the on-site energy for large angular momenta. The suppression of the higher harmonics is especially apparent in the bottom line of Fig.~\ref{fig:4}, which shows the deformation at a very large magnetic field with \mbox{$\omega_c/\gamma = 10$}. 

The deformation~\eqref{eq:trialtomographic} has angular projection \mbox{$a_n, b_n \sim (-1)^{(n+1)/2} e^{-n^2\delta\theta^2/8}$} (up to a normalization factor), and the dominant angular modes involved in the deformation scale as 
\begin{align}
\bar{m} \sim 1/\delta \theta \sim (k\xi)^{1/3} .
\end{align}
We expect the crossover from tomographic to magnetohydrodynamic scaling to occur when the magnetic field  interferes with the dominant odd-mode damping. This happens when the cyclotron frequency matches the dominant odd-mode damping rate, i.e., $\gamma' \bar{m}^4 \simeq \omega_c \bar{m}$, which gives Eq.~\eqref{eq:criticalfield}. In the hydrodynamic-tomographic transition region, where \mbox{$k\xi \simeq 1$}, this implies a large  cyclotron radius that is comparable to the odd-mode mean free path $v_F/\gamma_3$, in agreement with Eq.~\eqref{eq:sigmaTexpansion}. The critical field increases in the tomographic scaling regime, where it is however still significantly smaller than the magnetic field strength $\omega_c \simeq \gamma$ over which conventional hydrodynamic effects are suppressed. If the tomographic regime applies, we therefore expect a two-step scaling with increasing magnetic field, with a rapid suppression of tomographic transport at small fields of order $\omega_c^{\rm crit}$ to a magnetohydrodynamic form, followed by a suppression of hydrodynamics at much larger fields \mbox{$\omega_c \simeq \gamma$}.

\section{Kinetic description of tomographic magnetotransport}\label{sec:kinetic}

This section contains a detailed account and derivation of our main results that were summarized in the previous section. We begin in Sec.~\ref{sec:currentexpectation} with a discussion of the current expectation value. This is followed in Sec.~\ref{sec:variational} by a detailed account of the zero field structure of the Fermi surface along the hydrodynamic-tomographic-collisionless crossover, with variational bounds on the conductivity. Finally, Sec.~\ref{sec:derivativeexpansion} presents a derivative expansion that provides exact results at long wavelengths.

\subsection{Charge current expectation value}\label{sec:currentexpectation}

Within Fermi liquid theory, the charge current is defined as~\cite{pines18}
\begin{align}
{\bf j} &= - e N_0 \langle {\bf v} | h \rangle = - \frac{e v_F N_0}{2} \begin{pmatrix}
        h_{+1} + h_{-1} \\[1ex]
        i h_{+1} - i h_{-1}
    \end{pmatrix} , \label{eq:currentexpectation}
\end{align}
where \mbox{$N_0 = m^*/\pi\hbar^2 = 2n/(m^* v_F^2)$} is the density of states of the spinfull electron gas at the Fermi surface (with $n$ its density). [Note that in the following discussion we neglect the Landau parameters, which would otherwise add an overall factor \mbox{$(1+F_1)$} in Eq.~\eqref{eq:currentexpectation} for the renormalized velocity of the two-dimensional Fermi liquid.] The first expression in Eq.~\eqref{eq:currentexpectation} uses a scalar-product notation with
\begin{align}
\langle f | g \rangle &= \int_0^{2\pi} \frac{d\theta}{2\pi} f^*(\theta) g(\theta) , \label{eq:dotproduct}
\end{align}
and the second equality uses the harmonic basis~\eqref{eq:hangular}, with $h_{\pm 1}$ a solution of Eq.~\eqref{eq:tightbindingintro} as described in the introduction. The conductivity then reads
\begin{align}
\sigma_{ij} &= e^2 N_0 \, \langle v_i | G | v_j \rangle , \label{eq:conductivity}
\end{align}
where $G^{-1}$ is the linearized (static) Boltzmann operator, cf.~Eq.~\eqref{eq:fullkinetic},
\begin{align}
G^{-1} &= - L + \Bigl\{i {\bf k} \cdot {\bf v}(\theta) + \omega_c \frac{\partial}{\partial \theta}\Bigr\} , \label{eq:greensfunction}
\end{align}
which determines the quasiparticle deformation by
\begin{align}
    G^{-1} h &= - {\bf v}(\theta) \cdot e {\bf E} . \label{eq:currentshort}
\end{align}
Note that the expression in Eq.~\eqref{eq:conductivity} can be interpreted as a Green's function that describes the propagation from an initial site $m=\pm1$ (for the velocity component $v_j$) to a final site $m=\pm1$ (for the velocity component $v_i$). 

As discussed in Sec.~\ref{sec:mainresults}, to determine the different components of the conductivity tensor, we choose ${\bf E}$ along the $y$ direction (i.e., specify \mbox{$\theta_E = \pi/2$}) and  ${\bf k}$ as before along the $x$ direction. In this setup, the different components of the conductivity tensor follow as \mbox{$\sigma_L = \sigma_{xx}$}, \mbox{$\sigma_T = \sigma_{yy}$}, and \mbox{$\sigma_H = \sigma_{xy} = -\sigma_{yx}$}. We determine the conductivity numerically from the solution of Eq.~\eqref{eq:currentshort} described in Appendix~\ref{sec:harmonicprojection} and Refs.~\cite{hofmann22,setiawan22}.

\subsection{Variational bounds on the conductivity}\label{sec:variational}

For a general variational trial deformation $h(\theta)$ that is not necessarily a solution of Eq.~\eqref{eq:currentshort}, a rigorous lower bound for the conductivity can be derived, which reads
\begin{align}
\sigma_T \geq e^2 N_0 \, \frac{|\langle v_y | h \rangle |^2}{\langle h | G^{-1} | h\rangle} , \label{eq:conductivitybound}
\end{align} which follows from an application of the Cauchy-Schwarz inequality, where the equality holds if $h$ solves Eq.~\eqref{eq:currentshort}. 
A derivation of this expression is summarized in Appendix~\ref{sec:cauchyschwarz}. Note that the overall normalisation of $h$ does not enter in Eq.~\eqref{eq:conductivitybound}.

\begin{figure}
\includegraphics{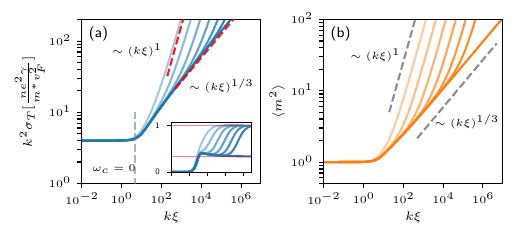}\\[-2ex]
${}\hspace{1.05cm}$\includegraphics{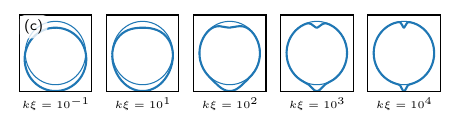}
\caption{
(a) Transverse conductivity at zero magnetic field as a function of dimensionless momentum $k\xi$ with power-law damping $\gamma' m^4$ (blue line) and for the odd-parity damping~\eqref{eq:odddamping} with six different ratios of the even- to odd-rates $\gamma'/\gamma = 10^{-6}, 10^{-5}, 10^{-4}, 10^{-3},10^{-2}$ (ordered by increasing opacity). The red dashed lines are the variational lower bounds~\eqref{eq:sigmatomographic} and~\eqref{eq:sigmacolless}, and the vertical gray dashed line marks the hydrodynamic-tomographic crossover \mbox{$k\xi={\it O}(1)$}. The inset shows the logarithmic derivative of the conductivity with respect to momentum, which reveals the local scaling exponent with wave number. (b) Variance $\langle m^2 \rangle$ of the Fourier modes of the Fermi surface deformation, for the same parameter values as in panel (a). (c) Fermi surface deformation in momentum space at $kx=0$ for five different momenta \mbox{$k\xi = 10^{-1}, 10^1, 10^2, 10^3$}, and $10^4$.
 }
\label{fig:5}
\end{figure}

It is instructive to discuss in more detail the transverse conductivity and the Fermi surface deformation involved in the electromagnetic response in the absence of a magnetic field (\mbox{$\omega_c=0$}) and impurity scattering ($\gamma_i=0$). Such a reduced model shows a direct transition from a hydrodynamic limit at \mbox{$k\xi \ll 1$} to a tomographic limit at \mbox{$k\xi \gg 1$}, in which the conductivity diverges at large momenta without a separate collisionless regime (which in the presence of impurity scattering will cross over to an Ohmic regime). This is shown in Fig.~\ref{fig:5}(a), which plots the transverse conductivity as a function of $k\xi$ using the odd-parity damping function~\eqref{eq:odddamping} for a pure power-law damping $\gamma' m^4$ (solid continuous blue line) and for different ratios \mbox{$\gamma'/\gamma = 10^{-6}, 10^{-5}, 10^{-4}, 10^{-3},10^{-2}$} with the odd-mode damping in Eq.~\eqref{eq:odddamping} (blue lines with increasing opacity). 

First, in the hydrodynamic limit, the Fermi surface deformation is a simple transverse current deformation with a small phase-shifted quadrupole admixture,
\begin{align}
    h_{\rm hydro}(\theta) &\sim \sin \theta - i \frac{v_Fk}{2\gamma} \sin 2 \theta , \label{eq:deformationhydro}
\end{align}
i.e., it involves only the odd-parity current components~\mbox{$m=\pm 1$} and the even-parity components~\mbox{$m=\pm 2$}. The relative magnitude of the even and odd components in Eq.~\eqref{eq:deformationhydro} follows from a solution of Eq.~\eqref{eq:currentshort}. The structure of this deformation is illustrated in Fig.~\ref{fig:5}(c), which shows the numerically determined Fermi surface [from Eq.~\eqref{eq:currentshort}] at \mbox{${\bf r} = 0$} with a pure power-law damping $\gamma' m^4$ for selected momenta along the hydrodynamic to tomographic crossover. The left figure shows a hydrodynamic deformation, where the transverse current component displaces the center of mass vertically. The quadrupole deformation in Eq.~\eqref{eq:deformationhydro} with a relative imaginary factor introduces a small (parity-symmetric) ellipticity of the Fermi surface at a distance \mbox{$r = 2\pi/q$} (not shown here). The hydrodynamic limit of the transverse conductivity corresponding to Eq.~\eqref{eq:deformationhydro} is given by 
\begin{align}
k^2 \sigma_{\rm hydro}(k) &= \frac{4 n e^2 \gamma}{m^* v_F^2} ,
\end{align}
which also follows from the variational form~\eqref{eq:conductivitybound}. The hydrodynamic Fermi surface deformation~\eqref{eq:deformationhydro} does not include higher odd-parity modes and is thus independent of the odd-mode damping.

Second, for sufficiently large separation \mbox{$\gamma' \ll \gamma$},  there is an intermediate tomographic scaling regime. By dimensional analysis, this regime is expected to start for momenta exceeding \mbox{$k\xi=1$}. Here, the third harmonic component starts to contribute, and the onset is independent of $\gamma/\gamma'$, which controls the damping of larger $m$ modes. (Indeed, the criterion \mbox{$k = v_F/\sqrt{\gamma\gamma_3}$} yields \mbox{$k\xi = 3\sqrt{3} \approx 5.196$}, in quite accurate agreement with the point in Fig.~\ref{fig:5} at which the conductivity starts to deviate from the hydrodynamic value.) At larger momenta, the Fermi surface deformation becomes strongly peaked in a narrow angle interval along the transverse direction (cf.~the middle and right plots in Fig.~\ref{fig:5}). We model this by the normalized trial deformation~\eqref{eq:trialtomographic} which is spread out over an angular width $\delta \theta$ given in Eq.~\eqref{eq:widthtomographic}. The corresponding lower bound on the transverse conductivity reads
\begin{align}
k^2 \sigma_{\rm tomo}(k) &= \frac{n e^2 \gamma}{m^* v_F^2} \times \frac{20}{3} \biggl(\sqrt{\frac{2}{15}}\frac{k\xi}{\pi^{3/2}}\biggr)^{1/3} . \label{eq:sigmatomographic}
\end{align}
The scaling with $k\xi$ agrees with a calculation by~\textcite{ledwith19} from a variational argument applied to the numerator of Eq.~\eqref{eq:conductivitybound}, and is the basis of the anomalous temperature scaling discussed by~\textcite{zeng24}. Here, we are able to obtain a rigorous interpretation as a lower bound and provide the prefactor of this estimate. The result~\eqref{eq:sigmatomographic} is shown in Fig.~\ref{fig:5} as a dashed line. The lower bound agrees remarkably well with the full solution, with a maximal deviation of $9.6\%$ from the full result.

Third, at even larger momenta, there is a crossover from a tomographic to a conventional collisionless regime with a different power-law scaling \mbox{$k^2\sigma(k) \sim k\xi$}. We attribute this crossover to a Fermi surface deformation that predominantly involves odd and even modes at such large $m$ that \mbox{$\gamma_m\sim \gamma$}, cf.~Eq.~\eqref{eq:odddamping}.  Repeating the same variational argument as above for $\gamma_{m\, {\rm odd}}=\gamma$ gives
\begin{align}
    \delta \theta_{\rm colless} &= \frac{2\gamma}{v_Fk} = \sqrt{\frac{\gamma}{\gamma'}} \frac{2}{k\xi} ,
\end{align}
which decays even faster than in the tomographic regime, Eq.~\eqref{eq:widthtomographic}.
The resulting lower bound on the transverse conductivity reads
\begin{align}
k^2 \sigma_{\rm colless}(k) &= \frac{n e^2 \gamma}{m^* v_F^2} \times \sqrt{\frac{2}{\pi}} \frac{2v_Fk}{\gamma} = \frac{n e^2 \gamma}{m^* v_F^2} \times \sqrt{\frac{8 \gamma'}{\pi \gamma}} \, k\xi  \label{eq:sigmacolless}
\end{align}
with the expected linear scaling in the wave number. Equation~\eqref{eq:sigmacolless} agrees well with the results shown in Fig.~\ref{fig:5}(a), this time with a maximal deviation of $20.2\%$ from the full result.

The inset of Fig.~\ref{fig:5}(a) shows the logarithmic derivative of the transverse conductivity with respect to wave number $k\xi$, which gives a local scaling exponent \mbox{$k^2\sigma(k) \sim (k\xi)^\beta$}. The inset clearly shows the crossover from the hydrodynamic regime (with \mbox{$\beta = 0$}) to the tomographic regime (with \mbox{$\beta = 1/3$}) and the collisionless regime (with \mbox{$\beta = 1$}) (cf.~also Fig.~\ref{fig:1}). For a larger odd mode damping coefficient $\gamma'$, the crossover from the tomographic to the collisionless regime occurs at increasingly smaller momenta. For the largest odd-mode damping coefficient \mbox{$\gamma'/\gamma \simeq 10^{-2}$} shown in Fig.~\ref{fig:5}(a) (light blue line), no separate tomographic scaling regime is visible.

We proceed to analyze the Fermi surface deformation involved in the transverse response in more detail. Figure~\ref{fig:5}(b) shows the width \mbox{$\sqrt{\langle m^2 \rangle}$} of the angular mode decomposition of $\delta f$ as a function of dimensionless wave number $k\xi$ for the same damping parameter values as in Fig.~\ref{fig:5}(a). As expected, in the hydrodynamic limit only the first few modes (\mbox{$m=0$} and \mbox{$m=\pm 1$} are involved in the response), with \mbox{$\sqrt{\langle m^2 \rangle} = 1$}. For an intermediate tomographic scaling, an increasing number of odd modes contribute with a width that scales as \mbox{$\sqrt{\langle m^2 \rangle} \sim (k\xi)^{1/3}$} [indicated by the dashed line in Fig.~\ref{fig:5}(b)]. This is consistent with the variational argument, by which we expect \mbox{$\sqrt{\langle m^2 \rangle} \sim 1/\delta \theta$}.  Likewise, in the collisionless regime, the width of the distribution scales as \mbox{$\sqrt{\langle m^2 \rangle} \sim (k\xi)^{1}$} [second gray dashed line in Fig.~\ref{fig:5}(b)], again consistent with the variational argument.

\subsection{Derivative expansion}\label{sec:derivativeexpansion}

In the long-wavelength hydrodynamic limit (i.e., for \mbox{$k \xi \ll 1$}), the conductivity is determined by the hydrodynamic modes \mbox{$m=0$} and \mbox{$m=\pm 1$} in Eq.~\eqref{eq:tightbindingintro}, with dissipative corrections set by the even-parity damping due to the coupling to the even-parity shear (quadrupole) modes \mbox{$m=\pm 2$}. Coupling to even higher modes is suppressed, being of higher powers of $k$. Taking into account modes with \mbox{$|m|\leq 2$} gives the leading-order transverse conductivity in Eq.~\eqref{eq:sigmaTexpansion}. 

The higher-order corrections in Eq.~\eqref{eq:sigmaTexpansion} are obtained in a derivative expansion that treats the off-diagonal (advective) term in the Green's function~\eqref{eq:greensfunction} as a perturbation. Formally, we separate \mbox{$G^{-1} = G_0^{-1} + J$}, 
where $G_0^{-1}$ includes the hydrodynamic block in the kinetic equation~\eqref{eq:kineticmatrix} as well as all diagonal terms for \mbox{$|m|>2$}, and $J$ contains all off-diagonal terms not included in $G_0^{-1}$. Explicitly,
\begin{align}
        G_0^{-1} = \begin{pmatrix}
        \ddots & \ddots &  &  &  &  &  \\[-1.5ex]
        \ddots & \epsilon_3 & 0 &  &  & &  \\[0.5ex]
         & 0 & \epsilon_2 & \frac{v_Fq}{2} &  &  &  0 &  \\[0.5ex]
         & & \frac{v_Fq}{2} & \epsilon_1 & \frac{v_Fq}{2} & &  &  \\[0.25ex]
         & &  & \frac{v_Fq}{2} & \epsilon_0 & \frac{v_Fq}{2} &  &  \\[0.25ex]
         & &  &  & \frac{v_Fq}{2} & \epsilon_{-1} & \frac{v_Fq}{2} &  \\[0.5ex]
         & & 0 &  &  & \frac{v_Fq}{2} & \epsilon_{-2} & 0 \\[-0.5ex]
         & &  &  &  &  & 0 & \epsilon_{-3} & \ddots \\
         & &  &  &  &  & & \ddots & \ddots
    \end{pmatrix} ,
\end{align}
where \mbox{$\epsilon_m = - (\omega - m \omega_c + i\gamma_m)$} (cf.~Appendix~\ref{sec:harmonicprojection}) and
\begin{align}
        J &= \frac{v_Fq}{2} \begin{pmatrix}
        \ddots & \ddots &  &  &  &  &  \\[-1.5ex]
        \ddots & 0 & 1 &  &  & &  \\[0.5ex]
         & 1 & 0 &  &  &  &  0 &  \\[-1ex]
         & &  &   &   & &  &  \\[-1.5ex]
         & &  &   & \ddots &   &  &  \\[-1ex]
         & &  &  &   &   &   &  \\[-1ex]
         & & 0 &  &  &   & 0 & 1 \\[-1ex]
         & &  &  &  &  & 1 & 0 & \ddots \\[-0.5ex]
         & &  &  &  &  & & \ddots & \ddots
    \end{pmatrix} .
\end{align}
Equation~\eqref{eq:greensfunction} is then expanded in a Lippmann-Schwinger form
\begin{align}
G &= G_0 + G_0 J G_0 + G_0 J G_0 J G_0 + \ldots , \label{eq:GExpansion}
\end{align}
In the tight-binding representation, the leading-order result~\eqref{eq:sigmaTexpansion} then follows as
\begin{align}
\sigma_T &= [G_0]_{1,-1} + [G_0]_{-1,1} - [G_0]_{11} - [G_0]_{-1,-1} ,
\end{align}
where the subindices $[G_0]_{m,m'}$ denote angular momentum matrix indices. This expression can be thought as propagation from lattice sites \mbox{$m=\pm1$} to sites \mbox{$m=\pm1$} via \mbox{$m=0$} and \mbox{$m=\pm2$} but not involving any other sites. Higher-derivative corrections then describe additional hopping steps outside this region and their return. For example, a term that contributes to the leading correction to the hydrodynamic result is
\begin{align}
[G_0 J G_0 J G_0]_{11} = [G_0]_{21} J_{32} [G_0]_{33} J_{23} [G_0]_{21} ,
\end{align}
which describes propagation from site $1$ to site $2$, followed by a hopping from $2$ to $3$, for which the propagator is diagonal, and then a return hopping from $3$ to $2$ and propagation back to site $1$. In full, this gives the transverse conductivity to second order in $k$ stated and discussed in Eq.~\eqref{eq:sigmaTexpansion}. 

We remark that the long-wavelength expansion of the transverse conductivity in Eq.~\eqref{eq:sigmaTexpansion} can also be obtained from a direct asymptotic analysis of the linearized Boltzmann equation. Such a so-called ``Hilbert expansion'' gives a rigorous asymptotic solution to the Boltzmann equation in the near-hydrodynamic limit. This is achieved by first scaling the linearized Boltzmann equation, which reveals dimensionless parameters that quantify the flow regime; in particular, the Knudsen number, \mbox{$\text{Kn}= v_F k /\gamma$}, quantifies the relative prevalence of collisional to advective processes. The long-wavelength limit, where the flow is dominated by collisions, corresponds to the small Knudsen number limit, \mbox{$\text{Kn}\ll1$}. In this limit, the distribution function and each of its moments (e.g., the current density) can be expanded in powers of Knudsen number. Substituting these into the Boltzmann equation, collecting like powers of $\text{Kn}$, and solving the resulting equations then gives the transverse conductivity at each-order of expansion (see Appendix~\ref{app:HExpansion} for details). To second-order in the Knudsen number, the transverse conductivity is given by Eq.~\eqref{eq:sigmaTexpansion}. This provides an independent validation of the long-wavelength limit in Eq.~\eqref{eq:sigmaTexpansion}.

\section{Discussion and outlook}\label{sec:outlook}

In summary, we have quantified the magnetic-field induced suppression of tomographic electron transport, a recently conjectured electron transport regime that combines hydrodynamic transport of even-parity modes with ballistic transport of long-lived odd-parity excitations. The core hypothesis, which we illustrate with calculations of the static transverse conductivity, is that the crossover from tomographic to Ohmic transport proceeds in two steps as a function of increasing magnetic field: First, tomographic transport is suppressed at rather weak magnetic fields that correspond to a large cyclotron radius of the order of the dominant odd-parity mean free path, followed by a suppression of hydrodynamic transport of even-parity modes at even larger fields.

\begin{figure}
    \centering    \includegraphics[width=1\linewidth]{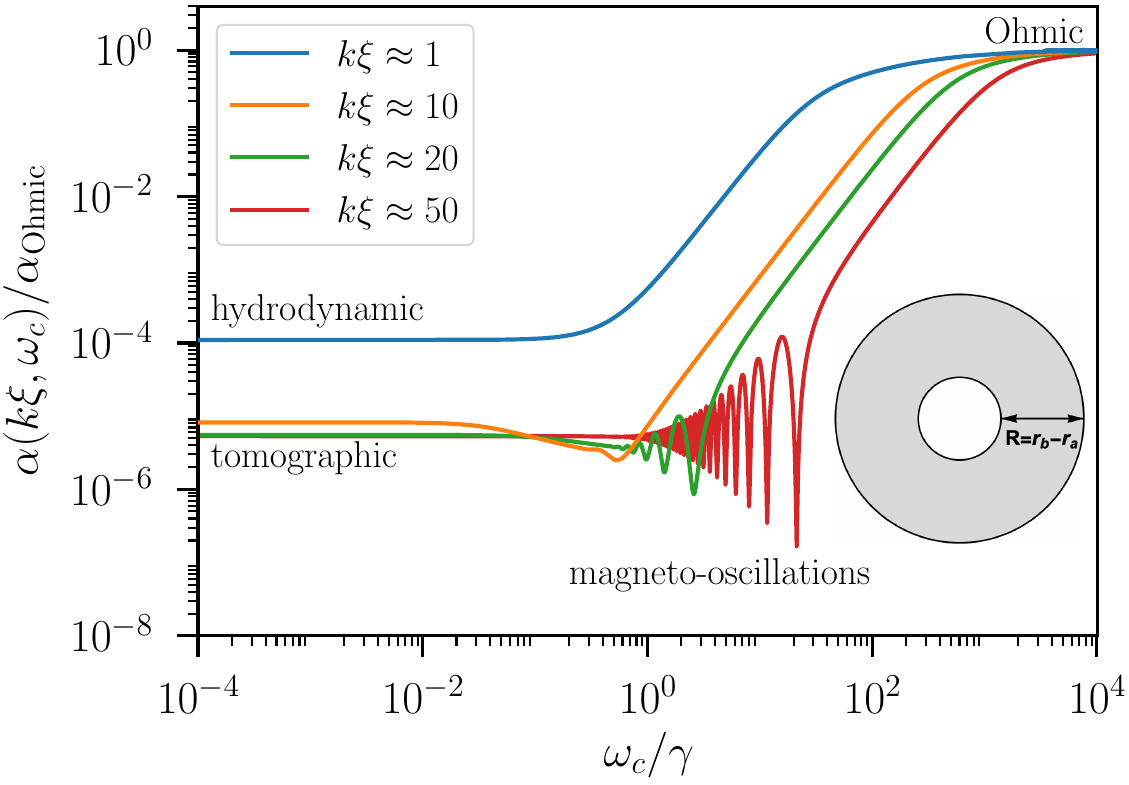}
    \caption{Magnetoresistance in a Corbino device (inset sketch) displaying the transition to an Ohmic regime with increasing magnetic field. Different values of $ k \xi $ with \mbox{$k=1/R$} are set by the system size, \mbox{$R / \ell_h = 100, 10, 5$}, and~$2$ in units of the even-parity mean free path \mbox{$\ell_h = v_F/\gamma$}, which correspond to the blue, orange, green, and red curves, respectively. The blue curve starts in the hydrodynamic regime at low fields, and other curves start in the tomographic regime. We use \mbox{$ \gamma_i/\gamma  = 10^{-7}$} and an odd-even scattering rate ratio of \mbox{$ \gamma' / \gamma = 10^{-4} $}. 
 }    \label{fig:6}
\end{figure}

As an outlook, let us briefly comment on observable signatures of a magnetic-field induced suppression of presumed tomographic scaling in the recent experiment by \textcite{zeng24}. The experiment measures the magnetic-field derivative of the electrical longitudinal magnetoresistance ${\cal R}(B)$ of both monolayer and bilayer graphene Corbino devices, 
\begin{align} 
\alpha(B) = \frac{d {\cal R}(B)}{d(B^2)} ,
\end{align} 
where this is understood as a local derivative at fixed magnetic background field. The factor $\alpha$ takes distinct constant values in the Ohmic and hydrodynamic regime, with \mbox{$\alpha_{\rm Ohmic}= \ln(r_b/r_a)/(2\pi \rho_o (ne)^2) > \alpha_{\rm hydro}$}, where $r_a$ and $r_b$ are the inner and outer radii of the Corbino disk (cf.~the inset sketch in Fig.~\ref{fig:6}) and we define the Ohmic resistivity \mbox{$ \rho_o = {m^\ast \gamma_i}/{ne^2} $}, which allows to disentangle electron hydrodynamic and Ohmic effects by measuring $\alpha$ at small and large magnetic fields. This interpretation is borne out of a solution of the Stokes-Ohm equation for electron flow in a Corbino geometry, which incorporates both viscous and Ohmic dissipation,
\begin{align} \label{eq:Stokes-Ohm}
{\bf j} - \frac{\eta}{(e n)^2 \rho_o} \nabla^2 {\bf j} =  \frac{{\bf E}}{\rho_o} -\frac{1}{en \rho_o} {\bf j} \times {\bf B} .
\end{align}
A comparison between experimental measurements of $\alpha$ and the Stokes-Ohm prediction provides an estimate for $\eta$ and thus a viscous transport lifetime $\tau_\eta$, which at zero magnetic field shows the previously discussed anomalous linear temperature scaling as opposed to a quadratic Fermi-liquid scaling. Although the Stokes-Ohm equation is strictly applicable only in the hydrodynamic regime, Refs.~\cite{zeng24,ledwith19} suggest to extend its domain of validity to the tomographic and collisionless regime by replacing the viscosity by a wave-vector dependent form, $\eta(k)$, where \mbox{$k=1/R$} is linked to the inverse characteristic device dimension \mbox{$R = r_b - r_a$}. Formally, following the hydrodynamic definition of the viscosity, this amounts to replacing $\eta$ by $e^2n^2/[k^2 \sigma_T(k,B)]$, which depends on both $k$ and $B$. Indeed, at zero magnetic field, the power-law scaling with wave number $k$ then translates to different power laws in both density and temperature, which are \mbox{$\tau_\eta = n^{1/6}/T$} in the tomographic~\cite{kryhin23,zeng24}, \mbox{$\tau_\eta = 1/n$} in the Ohmic, \mbox{$\tau_\eta = n/T^2$} in the hydrodynamic, and \mbox{$\tau_\eta = 1/\sqrt{n}$} in the collisionless limit.

Figure~\ref{fig:6} shows the magnetoresistance factor $\alpha$ as determined from a solution of the Stokes-Ohm equation~\eqref{eq:Stokes-Ohm}~\cite{zeng24} with a transverse conductivity taken from our calculations as a function of the cyclotron frequency $\omega_c$ for different values of $k\xi$. We assume the same parameters as in Fig.~\ref{fig:5} with an odd-even effect in the quasiparticle relaxation rates with \mbox{$\gamma'/\gamma = 10^{-4}$} and a small impurity scattering rate \mbox{$\gamma_i/\gamma=10^{-7}$}. Different values of \mbox{$k\xi$}, which determine the  expected transport regime at zero magnetic field, are obtained by varying the system size \mbox{$R$}. Consistent with Fig.~\ref{fig:1}, for small values of \mbox{$k\xi \lesssim 1$} (blue curve), there is a direct transition from a hydrodynamic small-field regime to an Ohmic high-field regime without pronounced features at intermediate fields. This smooth profile appears in good agreement with recent measurements (cf.~Fig.~1(c) in Ref.~\cite{zeng24}). For larger values \mbox{$1\lesssim k\xi<(\gamma/\gamma')^{3/4}$}, the low-field magnetoresistance is in the tomographic scaling region (orange, green, and red curves). The analysis in Fig.~\ref{fig:6} reveals interesting features that are absent in the conventional hydrodynamic regime and that could be used to probe the tomographic state of a two-dimensional electronic Fermi liquid in future experiments. The suppression of tomographic transport at low fields is apparent by an additional minimum of the resistance at intermediate fields (orange and green lines) for larger discs, with magneto-oscillations regimes at higher fields for smaller discs (green and red curves), before the resistance rises to the Ohmic value at high fields. In particular, the resistance minimum occurs at very small magnetic fields and could be used to determine the odd-parity decay rate $\gamma'$. We conclude by briefly discussing the effect of disorder on this picture: Since the direct momentum-relaxing term in the Stokes-Ohm equation~\eqref{eq:Stokes-Ohm} only separates the effect of impurity scattering on the hydrodynamic modes \mbox{$m=\pm 1$}, the electronic contribution beyond the hydrodynamic limit (i.e., the conductivity at finite wave number) is still affected: An impurity scattering rate $\gamma_i$ will induce a constant shift \mbox{$\gamma_{m \geq 2} \to \gamma_{m \geq 2} + \gamma_i$}, and significant impurity scattering \mbox{$\gamma_i = {\it O}(\gamma)$} will also suppress tomographic scaling.

{\it Note added:} Recently, an exciting study appeared that reports possible signatures of odd-even lifetimes in the width of cyclotron resonances in graphene photoconductivity~\cite{moiseenko24}. Compared to our analysis, which discusses the transverse static conductivity, this work considers features of the optical magnetoresponse, i.e., the longitudinal conductivity at finite frequency.

\begin{acknowledgments}
J.H. thanks Cory Dean, Ulf Gran, Jeff Maki, Seth Musser, and Sankar Das Sarma for discussions. This work is supported by Vetenskapsr\aa det (Grant No. 2020-04239 and 2021-03685) and Nordita.
\end{acknowledgments}

\appendix

\section{Kinetic equation in a magnetic field}\label{sec:introkinetic}

This Appendix contains a derivation of the kinetic equations in a magnetic field in Sec.~\ref{sec:introkinetic}. Classical papers are Refs.~\cite{lee75,simon93}, and a textbook discussion can be found in Ref.~\cite{pines18}. 

In the absence of external perturbations, quasiparticles are in global equilibrium described by the Fermi-Dirac distribution
\begin{align}
f_0({\bf p}) = \frac{1}{e^{\beta (\varepsilon({\bf p}) - \mu)} + 1} ,
\end{align}
which depends on the bare band dispersion $\varepsilon({\bf p})$ that is independent of position. Quasiparticle excitations induce a deviation from global equilibrium,
\begin{align}
\delta f(t,{\bf r},{\bf p}) &= f(t,{\bf r},{\bf p}) - f_0({\bf p}) , \label{eq:deviationglobal}
\end{align}
which induces a mean-field shift in the single-particle energy, \begin{align}
\tilde{\varepsilon}(t, {\bf r}, {\bf p}) = \varepsilon({\bf p}) + \delta \varepsilon(t, {\bf r}, {\bf p}) , \label{eq:qpenergy}
\end{align}
where $\delta \varepsilon(t, {\bf r}, {\bf p})$ is the Fermi-liquid mean-field correction
\begin{align}
\delta \varepsilon(t, {\bf r}, {\bf p}) &= \sum_{\bf q} F({\bf p}, {\bf q}) \delta f(t, {\bf r}, {\bf q})
\end{align}
with $F({\bf p}, {\bf q})$ parametrizing the interacting energy between quasiparticles. 

The time evolution of the distribution function $f(t,{\bf r},{\bf p})$ in response to an external force is governed by a kinetic equation
\begin{align}
&\biggl\{\frac{\partial}{\partial t} + \dot{\bf r} \cdot \frac{\partial}{\partial {\bf r}}  + \dot{\bf p} \cdot \frac{\partial}{\partial {\bf p}} \biggr\} f(t,{\bf r},{\bf p}) = I[\delta \bar{f}] , \label{eq:kinetic}
\end{align}
where $I[\delta \bar{f}]$ is the collision integral, $\delta \bar{f}$ the deviation from local equilibrium (see below), and we use the semiclassical equations of motion for a Bloch state in a magnetic and electric field ${\bf B}$ and ${\bf E}$~\cite{pines18},
\begin{align}
\dot{\bf r} &= 
\tilde{\bf v}(t, {\bf r}, {\bf p}) \label{eq:dotr} \\[1ex]
\dot{\bf p} &= 
\biggl(- e {\bf E}(t, {\bf r}) - \frac{\partial \tilde{\varepsilon}(t, {\bf r}, {\bf p})}{\partial {\bf r}}\biggr) 
- \tilde{\bf v}(t, {\bf r}, {\bf p}) \times e{\bf B} \label{eq:dotp}
\end{align}
with \mbox{$e>0$} and band velocity \mbox{$\tilde{\bf v}(t, {\bf r}, {\bf p}) = {\partial \tilde{\varepsilon}(t, {\bf r}, {\bf p})}/{\partial {\bf p}}$}.
Within this notation, ${\bf p}$ is the (kinematic) momentum, not wave number~\cite{pines18}. The right-hand side of Eq.~\eqref{eq:kinetic} accounts for a change in the distribution function by collisions. These collisions do not relax the distribution to global equilibrium but to a local equilibrium distribution,
\begin{align}
\bar{f}_0(t,{\bf r},{\bf p}) = \frac{1}{e^{\beta (\tilde{\varepsilon}(t, {\bf r}, {\bf p}) - \mu 
)} + 1} ,
\end{align}
which by Eq.~\eqref{eq:qpenergy} implies \mbox{$ \bar f_0 -f_0 \approx (\partial f_0/\partial \varepsilon)\delta\varepsilon$}. The collision integral will thus be a  function of the deviation from local equilibrium 
\begin{align}
    \delta \bar{f}(t,{\bf r},{\bf p}) = 
f(t,{\bf r},{\bf p}) -\bar{f}_0(t,{\bf r},{\bf p}) ,
\end{align}
which is written in terms of the deviation from the global equilibrium \mbox{$\delta f(t,{\bf r},{\bf p})$}, Eq.~\eqref{eq:deviationglobal}, as
\begin{align}
\delta \bar{f}(t,{\bf r},{\bf p}) 
=   \delta f(t,{\bf r},{\bf p}) + \biggl(- \frac{\partial f_0}{\partial \varepsilon}\biggr) 
\delta \varepsilon(t, {\bf r}, {\bf p})
.
\end{align}
For a plane-wave perturbation, Eq.~\eqref{eq:planewave}, we parametrize the quasiparticle distribution function as
\begin{align}
\delta \bar{f}(t, {\bf r}, {\bf p}) &= e^{- i \omega t + i {\bf k} \cdot {\bf r}} \biggl(- \frac{\partial f_0}{\partial \varepsilon}\biggr) \bar{h}(\theta) , \label{eq:deformationhbar}
\end{align}
and likewise for $\delta f$ as stated in Eq.~\eqref{eq:deformation}, with a deformation function $h(\theta)$ [or $\bar{h}(\theta)$, respectively] that depends on polar angle $\theta$ of ${\bf p}$ (not its magnitude) in momentum space. As discussed, $h(\theta)$ represents a rigid deformation of the Fermi surface that, in this parametrization, has units of energy or chemical potential. Note that at low temperature, we have \mbox{$(-{\partial f_0}/{\partial \varepsilon}) = \delta(\varepsilon - \mu)$}, which restricts the dynamics to an angular degrees of freedom at the Fermi surface. With this parametrization and defining the linearized collision integral
\begin{align}
    L[\bar{h}(\theta)] &= \frac{I[\delta \bar{f}]}{(- \frac{\partial f_0}{\partial \varepsilon})} , \label{eq:linearizedcollision}
\end{align}
the linearized kinetic Eq.~\eqref{eq:kinetic} becomes~\cite{lee75,simon93} 
\begin{align} \label{eq:fullkinetic}
&-i \omega h(\theta) 
-
L[\bar{h}(\theta)] 
+
\biggl\{
i {\bf k} \cdot {\bf v}(\theta) + \omega_c \frac{\partial}{\partial \theta}\biggr\} \bar{h}(\theta) \nonumber \\[1ex]
&\qquad = 
- 
\bigl\{ {\bf v}(\theta) \cdot 
e {\bf E}  \bigr\} ,
\end{align}
where \mbox{${\bf v}(\theta) = v_F (\cos \theta, \sin \theta)$} is the band velocity.

\section{Angular harmonic tight-binding representation of the kinetic equation}\label{sec:harmonicprojection}

This Appendix discusses the projection of the kinetic equations onto angular harmonics stated in Eq.~\eqref{eq:tightbindingintro}, focusing on the role of the magnetic field. 

To solve the linearized kinetic equation, we need to specify the form of the linearized collision integral~\eqref{eq:linearizedcollision}. By dint of the circular symmetry of the Fermi surface, the eigenfunctions of this linear operator are labeled by an angular momentum index $m$. We thus expand $h(\theta)$ [or $\bar{h}(\theta)$] in angular harmonics as stated in Eq.~\eqref{eq:hangular}, for which we can use a generalized relaxation-time approximation for the collision integral,
\begin{align}
    L[\bar{h}_m] = - \gamma_m \bar{h}_m ,
\end{align}
where $\gamma_m$ describes the relaxation rate of the particular deformation with angular index $m$. In this section, we show how the kinetic equation~\eqref{eq:fullkinetic} reduces to an effective tight-binding form in this angular basis that is easy to solve. Formally, this is done by projecting onto the harmonic basis using the scalar product~\eqref{eq:dotproduct}.

\begin{figure}
    \centering
\includegraphics[width=0.95\linewidth]{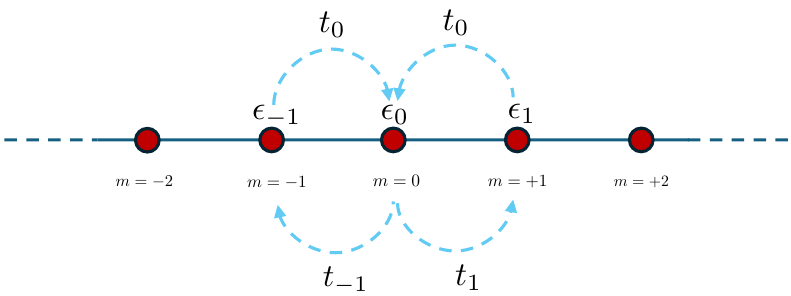}
    \caption{Illustration of the tight-binding structure of the low-temperature Fermi-liquid kinetic equation~\eqref{eq:tightbinding}. The angular mode index $m$ labels fictitious lattice positions with on-site energy $-\omega$ and damping $-i\gamma_m$. The magnetic field term, which is a linear function of $m$, plays the role of an effective electric field. The advective term couples adjacent harmonic modes and thus corresponds to a hopping term between sites with hopping amplitude $v_Fk/2$.}
    \label{fig:7}
\end{figure}

First, expanding the Landau function in the angular basis, we find
\begin{align}
N_0 F({\bf p}, {\bf q}) &= \sum_m e^{-im(\theta_{\bf p} - \theta_{\bf q})} F_m ,
\end{align}
where $F_m$ are the Landau parameters. This then gives a simple link between the harmonic components of the deviation from local and global equilibrium, 
\begin{align}
\bar{h}_m = \alpha_m h_m ,
\end{align}
with \mbox{$\alpha_m = 1 + F_m$}. In the following, we align the wave number of the perturbation along the $x$ direction, \mbox{${\bf k} = k \hat{\bf e}_x$} and choose a variable angle $\theta_E$ of the electric field  with the $x$ axis, \mbox{${\bf E}= E (\cos \theta_E, \sin \theta_E)$}. The angular momentum projection of Eq.~\eqref{eq:fullkinetic} then takes a simple form similar to one-dimensional tight-binding problem, where the angular index $m$ denotes a lattice site: 
\begin{align}\label{eq:tightbinding}
 \epsilon_m h_m + t_{m+1} h_{m+1}+t_{m-1} h_{m-1} = s_m .
\end{align}
Here, the on-site term $(\epsilon_m)$, hopping term $(t_{m\pm1})$, and source term $(s_m)$ are given by
\begin{align}
    \epsilon_m (\omega)&= - [\omega - m \omega_c\alpha_m + i\gamma_m\alpha_m] ,
    \\
    t_{m} (k)&=t_{-m} (k)= \frac{v_F k}{2} \alpha_{m}, 
    \\
    s_m &= i \frac{v_F e E}{2} \bigl [ e^{- i \theta_E}\delta_{m,1} + e^{+ i \theta_E} \delta_{m,-1}\bigr ].
\end{align}
The structure of this effective model is illustrated in Fig.~\ref{fig:7}: $\epsilon_m$ is an on-site energy, where the frequency induces a uniform energy shift $-\omega$ with an on-site dissipative term of strength $\gamma_m$. The magnetic field term is a linear function of $m$ and can be thought of as an effective constant electric field in mode space. The advective term $t_m$ induces a hopping between neighboring sites, which is nonreciprocal if the Landau parameters $F_m$ on these sites are different. The source terms $s_m$ do not have a direct interpretation in terms of a tight-binding Hamiltonian.

In matrix form, the effective tight-binding model~\eqref{eq:tightbinding} reads 
\begin{align}
    \begin{pmatrix}
        \ddots & \ddots &  &  &  &  &  \\[-1.5ex]
        \ddots & \epsilon_2 & t_{1} & 0 & 0 & 0 &  \\[0.25ex]
         & t_{2} & \epsilon_1 & t_0 & 0 & 0 &  \\[0.25ex]
         & 0 & t_{1} & \epsilon_0 & t_{-1} & 0 &  \\[0.25ex]
         & 0 & 0 & t_0 & \epsilon_{-1} & t_{-2} &  \\[-1.25ex]
         & 0 & 0 & 0 & t_{-1} & \epsilon_{-2} & \ddots \\
         &  &  &  &  & \ddots & \ddots
    \end{pmatrix} \begin{pmatrix}
        \vdots \\ h_{2} \\ h_{1} \\ h_0 \\ h_{-1} \\ h_{-2} \\ \vdots
    \end{pmatrix}
    = \begin{pmatrix}
        \vdots \\ 0 \\ s_{1} \\ 0 \\ s_{-1} \\ 0 \\ \vdots
    \end{pmatrix} . \label{eq:explicitmatrix}
\end{align} 
Projecting this linear set of equations onto the hydrodynamic modes (i.e., the $m = 0, \pm 1$ harmonics) gives the following expression:
\begin{align}
    \hat {\cal K}(\omega,k)  \begin{pmatrix}
        h_1 \\ h_0 \\ h_{-1}
    \end{pmatrix} =  \begin{pmatrix}
        s_1 \\ 0 \\ s_{-1}
    \end{pmatrix} , \label{eq:kineticmatrix}
\end{align}
with an effective Hamiltonian matrix 
\begin{align}
     \hat {\cal K}(\omega,k)  &= \begin{pmatrix}
         \epsilon_1 (\omega) & t_{0}(k) & 0
         \\[5pt]
         t_{1}(k) & \epsilon_0 (\omega)& t_{1}(k)
         \\[5pt]
         0 & t_{0} & \epsilon_{-1}(\omega) 
     \end{pmatrix}
     \nonumber\\[1ex]&
     +t_{2}(k) 
     \begin{pmatrix}
          x_2(\omega,k) & 0 & 0
         \\[5pt]
        0 & 0& 0
         \\[5pt]
         0 & 0 & x_{-2}(\omega,k)
     \end{pmatrix}.
\end{align}
Here, the coupling to higher harmonics \mbox{$|m|\geq 2$} is accounted for by the second term that contains the coefficient 
\begin{align}
x_m = \begin{cases} 
{h_m}/{h_{m-1}} & \text{if } m\ge 2 , \\[1ex]
{h_m}/{h_{m+1}} & \text{if } m \le -2 .
\end{cases} . \label{eq:defxm}
\end{align}
Since $ s_{|m|\ge 2} = 0 $, these coefficients solve a homogeneous tight-binding model
\begin{align}
 t_{m+1}  x_{m+1} + \frac{t_{m-1}}{x_{m}} +\epsilon_m = 0 ,
\end{align}
which is solved in terms of a continued fraction representation for $x_m$, which includes the effects of a magnetic field and general Fermi liquid parameters,
\begin{align}
x_{m}(\omega,k)= -\frac{t_{m-1}(k)}{\epsilon_m (\omega)+t_{m+1} (k) x_{m+1} (\omega,k)}, 
\end{align}
where here we indicate the dependence on frequency and wave number of the external perturbation explicitly. This continued fraction can be summed exactly using a modified Lenz algorithm, which is described in detail in Refs.~\cite{hofmann22,setiawan22}. This avoids a direct solution of the linear system of equations~\eqref{eq:tightbinding} or~\eqref{eq:explicitmatrix} by matrix inversion, which requires a cutoff in the angular index $m$ (i.e., a fixed matrix size). Such an explicit cutoff converges only slowly to the exact results and furthermore becomes inaccurate at large momenta when the Fermi surface deformation contains higher angular modes, which is exactly the regime we are predominantly interested in. Note that in the absence of a magnetic field, the coefficients $x_m$ and $x_{-m}$ are identical but they differ once the magnetic field is included. The solution for the coefficients $x_{\pm 2}$ is then substituted back in Eq.~\eqref{eq:kineticmatrix} to obtain $h_0$ and $h_{\pm 1}$. Higher harmonics of the deviation function $h_m$ are then obtained from Eq.~\eqref{eq:defxm}.

\section{Lower bound on the diagonal components of the conductivity}\label{sec:cauchyschwarz}

In this Appendix, we derive the variational bound~\eqref{eq:conductivitybound} on the diagonal components of the conductivity from a Cauchy-Schwartz inequality. Consider the (by definition) nonnegative norm
\begin{align}
&\langle h - \lambda \tilde{h} | G^{-1} (h - \lambda \tilde{h}) \rangle \nonumber \\
&= \langle h | G^{-1} h \rangle - 2 \lambda \langle \tilde{h} | G^{-1} h \rangle + \lambda^2 \langle \tilde{h} | G^{-1} \tilde{h} \rangle \geq 0 ,
\end{align}
where $h$ is the solution of Eq.~\eqref{eq:currentshort} and $\tilde{h}$ the variational ansatz for the Fermi surface deformation. 
The minimum of the left-hand side is reached for \mbox{$\lambda = \langle \tilde{h} | G^{-1} h \rangle/\langle \tilde{h} | G^{-1} \tilde{h} \rangle$}, which implies
\begin{align}
 \langle h | G^{-1} h \rangle \langle \tilde{h} | G^{-1} \tilde{h} \rangle \geq \langle \tilde{h} | G^{-1} h \rangle^2 .
\end{align}
Since $\sigma_{ii} = \langle v_i | G v_i \rangle = \langle h | G^{-1} h \rangle$, this Cauchy-Schwarz inequality implies the lower bound~\eqref{eq:conductivitybound}.

\section{Basis transformation at high momenta} \label{app:basis_high}

In this Appendix, we summarize the diagonalization of the Fermi liquid equations for large external wave numbers. As discussed in the main text, at such high momenta the Fermi surface deformation is strongly peaked on the Fermi surface and involves many angular harmonics. To determine the asymptotic high-field scaling, we do not need to exclude the hydrodynamic modes at \mbox{$m=0,\pm 1$} or Landau parameters at small $m$. Here we consider a simple odd-even staggered damping,
\begin{align}
\gamma_m &= \begin{cases}
\gamma & m \ {\rm even} , \\[1ex]
\bar{\gamma} & m \ {\rm odd} ,
\end{cases}
\end{align}
which generalizes the constant relaxation model~\eqref{eq:odddamping} for large $m$. The analysis applies for \mbox{$kr_c \gg 1$} and any value of the odd- and even-mode damping.

The starting point is to eliminate the advective term in the kinetic equation by measuring position with respect to the quasiparticle position on the Fermi surface~\cite{lee75,simon93}
\begin{align}
g(\theta) &= e^{i {\bf k} \cdot ({\bf r} - {\bf R}(\theta))} h(\theta) , \label{eq:modifiedexpansion}
\end{align}
with
\begin{align}
{\bf R}(\theta) &= - \frac{1}{\omega_c} \int_0^\theta d\phi \, {\bf v}(\phi) = - \frac{v_F}{\omega_c} \begin{pmatrix} \sin\theta \\ 1-\cos\theta \end{pmatrix} ,
\end{align}
which is a periodic function of $\theta$. In particular, for ${\bf k}$ in the $x$ direction, we have the Jacobi-Anger expansion
\begin{align}
e^{i {\bf k} \cdot {\bf R}(\theta)} &= e^{- i \frac{v_F k}{\omega_c} \sin \theta} = \sum_{m = -\infty}^\infty J_m(- X) e^{i m \theta} , \label{eq:highmomentumtrafo}
\end{align}
with \mbox{$X = v_F k/\omega_c$}, which obeys the identity \mbox{$i {\bf v} \cdot {\bf k} = - \omega_c^{-1} e^{- i {\bf k} \cdot {\bf R}(\theta)} \partial_\theta e^{i {\bf k} \cdot {\bf R}(\theta)}$}, such that the kinetic equation~\eqref{eq:kinetic} becomes
\begin{align} 
&- e^{i {\bf k} \cdot {\bf R}(\theta)} {\cal L}[h(\theta)] - \omega_c \cdot \frac{\partial}{\partial \theta} \bigl[ e^{i {\bf k} \cdot {\bf R}(\theta)} h(\theta) \bigr] \nonumber \\[1ex]
&\quad = - \bigl[e^{i {\bf k} \cdot {\bf R}(\theta)} {\bf v}(\theta)\bigr] \cdot 
e {\bf E}  .
\end{align}
After expanding $g(\theta)$ in Fourier series
\begin{align}
g(\theta) = e^{i {\bf k} \cdot {\bf R}(\theta)} h(\theta) &= \sum_m g_m \, e^{i m \theta} , 
\end{align}
the kinetic equation~\eqref{eq:kinetic} becomes
\begin{align}
 &- \sum_{m''} \biggl[\frac{\gamma+\bar{\gamma}}{2} \delta_{m'',m} + \frac{\gamma-\bar{\gamma}}{2} (-1)^{m''} J_{m-m''}(2X) \biggr] g_{m''} \nonumber \\
 &\qquad + i m \omega_c g_m  = {\bf c}_m \cdot e {\bf E} , \label{eq:highmomentumkinetic}
\end{align}
where
\begin{align}
{\bf c}_m &= 
v_F \begin{pmatrix} \frac{m}{X} J_m(X) \\[1ex] (-i) \frac{d}{dX} J_m(X)\end{pmatrix} .
\end{align}
Equation~\eqref{eq:highmomentumkinetic} describes a modified tight-binding model with an all-to-all hopping term between sites. As a consistency check, in the long-wavelength limit \mbox{$X\to0$} where the phase change in Eq.~\eqref{eq:modifiedexpansion} is unity and the model reduces to the previous tight-binding form, only the term \mbox{$m''=m$} in Eq.~\eqref{eq:highmomentumkinetic} remains, such that the term in square brackets becomes $\gamma$ for even $m$ and $\gamma'$ for odd $m$, as required.

In the large-momentum limit, \mbox{$X\to \infty$}, the asymptotic form of the Bessel function,
\begin{align}
J_{m-m''}(2X) &\rightarrow \sqrt{\frac{1}{\pi X}} \cos \Bigl(2X - \frac{(m-m'')\pi}{2} - \frac{\pi}{4}\Bigr) ,
\end{align}
ensures that the all-to-all coupling terms vanishes, such that the model~\eqref{eq:highmomentumkinetic} is indeed diagonal with a transverse conductivity
\begin{align}
\sigma_T &= e^2 N_0 \sum_m \frac{|c_{m,y}|^2}{i m \omega_c - \frac{\gamma+\bar{\gamma}}{2}} ,
\end{align}
where the momentum dependence is encoded in the current matrix element in the numerator. This conductivity has been evaluated by~\textcite{simon93} in closed analytical form in a related model with finite frequency but without damping. Analytically continuing their result in the complex frequency plane to \mbox{$-i\omega \to (\gamma+\bar{\gamma})/2$} and expanding to leading asymptotic order in \mbox{$X=v_Fk/\omega_c$} gives Eq.~\eqref{eq:asymptoticmagnetic} in the main text.

\section{Hilbert expansion for the response to a transverse wave}\label{app:HExpansion}

The Hilbert expansion technique has been widely reported and utilized to study classical gas flows in the near-hydrodynamic regime (i.e., in the long-wavelength limit)~\citep{sone07,sone02}, and has recently been employed to study electron flows in this regime~\cite{benshachar25a,benshachar25b}. This analysis method gives asymptotic solutions to the Boltzmann equation in the absence of flow domain boundaries. In this Appendix, we employ this technique to calculate the transverse conductivity of electron flows in the near-hydrodynamic regime. This analysis complements the derivative expansion in Sec.~\ref{sec:derivativeexpansion}, and, as we shall show, gives identical results up to ${\it O}(k^2)$.

The Hilbert expansion is performed by first scaling the velocity \mbox{$\mathbf{v}$}, the wavevector \mbox{$\mathbf{k}$}, and the distribution function \mbox{$\bar{h}$} by \mbox{$v_F$}, \mbox{$k$}, and \mbox{$eE\gamma/(k^2v_F)$}, respectively, which reformulates the kinetic equation~\eqref{eq:fullkinetic} in terms of dimensionless variables [see Eq.~\eqref{eq:scaledBTE} below]. This gives rise to dimensionless parameters that govern the flow regime: The Knudsen number \mbox{$\text{Kn} = k v_F/\gamma$}; the scaled relaxation rate of the \mbox{$m$}th angular harmonic \mbox{$\gamma_m/\gamma$}; and the scaled cyclotron frequency \mbox{$\omega_c /\gamma$}. The dimensionless kinetic equation~\eqref{eq:fullkinetic} for a steady state then reads
\begin{align}
    &-\frac{1}{\text{Kn}}J[\bar{h}]+\left\{i\cos\theta-\frac{1}{\text{Kn}}\frac{\omega_c}{\gamma}  \frac{\partial}{\partial \theta}\right\}\bar{h} = -\text{Kn} \sin\theta, \label{eq:scaledBTE}
\end{align}
where for a transverse perturbation we have used \mbox{$\theta_E=\pi/2$}. The scaled collision operator \mbox{$J$} satisfies,
\begin{subequations}
\begin{align}
    J[1]=&0,\\ J[\cos\theta]=J[\sin\theta]=&0, \\
    J[\cos2\theta]=J[\sin2\theta]=&-1,\\
    J[\cos m\theta]=J[\sin m\theta]=&-\frac{\gamma_m}{\gamma}, \quad m\in\mathbb{Z}_{\geq 3} .
\end{align}
\end{subequations}
The long-wavelength limit is then expressed as the limit \mbox{$\text{Kn}\ll 1$}, while assuming \mbox{$\gamma_m/\gamma\sim \omega_c/\gamma\sim 1$}. The latter assumption, which states that the cyclotron frequency is comparable to the collision frequency, is implicitly assumed in the derivative expansion of Sec.~\ref{sec:derivativeexpansion}. 

The solution for the distribution function and each of its moments is obtained in this limit through a regular perturbative expansion in \mbox{$\text{Kn}$},
\begin{subequations}
\begin{align}
    \bar{h} = & \frac{1}{\text{Kn}}\delta\mu^{(-1)} + \bar{h}^{(0)} + \text{Kn} \, \bar{h}^{(1)} + \text{Kn}^2 \,\bar{h}^{(2)} + \dots , \\
    \delta \mu = & \frac{1}{\text{Kn}}\delta\mu^{(-1)} + \delta \mu^{(0)} + \text{Kn} \, \delta \mu^{(1)} + \text{Kn}^2 \,\delta \mu^{(2)} + \dots ,  \\
    \mathbf{j} = & \mathbf{j}^{(0)} + \text{Kn} \, \mathbf{j}^{(1)} + \text{Kn}^2 \,\mathbf{j}^{(2)} + \dots ,
\end{align}\label{eq:Hilbert_hExpansion}%
\end{subequations}
where the dimensionless macroscopic variables at each order of expansion are
\begin{align}
    \delta \mu^{(n)} = \langle 1| \bar{h}^{(n)} \rangle, \quad \mathbf{j}^{(n)} =  \langle \mathbf{v} |  \bar{h}^{(n)} \rangle , \label{eq:Hilbert_moments}
\end{align}
for \mbox{$n\in\mathbb{N}$}, which are scaled by \mbox{$E/k$} and \mbox{$\gamma n e^2 E/(v_F^2 k^2 m^*)$}, respectively. Substituting Eq.~\eqref{eq:Hilbert_hExpansion} into Eq.~\eqref{eq:scaledBTE} and collecting powers of \mbox{$\text{Kn}$} gives
\begin{subequations}
\begin{align}
    J[\bar{h}^{(0)}]+\frac{\omega_c}{\gamma}\frac{\partial \bar{h}^{(0)}}{\partial \theta}=& i\cos\theta \, \delta \mu^{(-1)} , \label{eq:Hilbert_h0}\\
    J[\bar{h}^{(1)}]+\frac{\omega_c}{\gamma}\frac{\partial \bar{h}^{(1)}}{\partial \theta}=& i\cos\theta \, \bar{h}^{(0)}, \\
    J[\bar{h}^{(2)}] + \frac{\omega_c}{\gamma}\frac{\partial \bar{h}^{(2)}}{\partial \theta}=& i\cos\theta \, \bar{h}^{(1)} +\sin\theta, \\
    J[\bar{h}^{(n)}] + \frac{\omega_c}{\gamma}\frac{\partial \bar{h}^{(n)}}{\partial \theta}=& i\cos\theta \, \bar{h}^{(n-1)} ,\quad n\geq 3.
\end{align}\label{eq:Hilbert_hn}%
\end{subequations}
The above equations are then solved sequentially by projecting each equation onto the basis functions defined by $\sin(m\theta)$ and $\cos(m\theta)$. This gives the distribution function at each order of the expansion, \mbox{$\bar{h}^{(n)}$}, in terms of the moments (macroscopic variables) \mbox{$\delta \mu^{(n-m)}$} and \mbox{$\mathbf{j}^{(n-m)}$} for \mbox{$0\leq m \leq n$}. 
In particular, solving Eq.~\eqref{eq:Hilbert_h0} for $\bar{h}^{(0)}$ (the leading-order distribution function in $\text{Kn}$) gives a local equilibrium
\begin{align}
    \bar{h}^{(0)} = \delta \mu^{(0)} + 2 \mathbf{v}\cdot \mathbf{j}^{(0)} .
\end{align}
This shows that in this limit, analogous to the long-wavelength limit, a local equilibrium is established, and as such the conductivity is given by the continuum Stokes-Ohm solution. At subsequent orders of the expansion in \mbox{$\text{Kn}$} (i.e., contributions that vanish in the long-wavelength limit \mbox{$\text{Kn}\to0$} and thus provide finite-wavelength corrections to the continuum theory), the distribution function attains higher-order moments. These arise due to the advection term [the \mbox{$i \cos\theta$} terms on the right hand side of Eq.~\eqref{eq:Hilbert_hn}], and the trigonometric identities
\begin{subequations}
\begin{align}
    \cos\theta \sin(m\theta)= & \frac{\sin((m+1)\theta)+\sin((m-1)\theta)}{2} , \\ 
    \cos\theta \cos(m\theta) = & \frac{\cos((m+1)\theta)+\cos((m-1)\theta)}{2} .
\end{align}
\end{subequations}
Hence, through application of the advection operator, modes of order \mbox{$m$} in the distribution function \mbox{$\bar{h}^{(n)}$} couple to modes of order \mbox{$m+1$} and \mbox{$m-1$} in the distribution function \mbox{$\bar{h}^{(n+1)}$}---this is analogous to the ``hopping'' term in the derivative expansion.

Substituting the solutions for \mbox{$\bar{h}^{(n)}$} into the moment definitions in Eq.~\eqref{eq:Hilbert_moments} then gives a set of equations for the macroscopic variables. In particular, the equations for the components of $\mathbf{j}^{(0)}$, $\mathbf{j}^{(1)}$, and $\mathbf{j}^{(2)}$, are
\begin{subequations}
\begin{align}
    j_x^{(n)} = & 0 , \qquad n\in\{0,1,2\}, \\
    j_y^{(0)} = & 4\biggl[1+\left(\frac{2\omega_c}{\gamma}\right)^2\biggr] , \\
    j_y^{(1)} = & 0 , \\
    j_y^{(2)} = & -\frac{\left(12\gamma \omega_c^2+4\gamma_3\omega_c^2-\gamma^2\gamma_3\right)\gamma }{4\left(\gamma_3^2+9\omega_c^2\right)\left(\gamma^2+4\omega_c^2\right)} j_y^{(0)} .
\end{align}\label{eq:Hilbert_govEqns}%
\end{subequations}
Sequential solution of the equations in Eq.~\eqref{eq:Hilbert_govEqns} gives the asymptotic solution to $j_y$ up to $O(\text{Kn}^2)$. Restoring units then produces the asymptotic transverse conductivity in Eq.~\eqref{eq:sigmaTexpansion}. This result is identical to that obtained via the derivative expansion in Sec.~\ref{sec:derivativeexpansion}.

\bibliography{bib_magnetotransport}

\end{document}